\documentclass[amsmath,superscriptaddress,twocolumn]{revtex4-2}

\setcitestyle{super}
\usepackage{siunitx}
\usepackage{graphicx}
\usepackage[colorlinks=true,citecolor=blue,urlcolor=blue,linkcolor=blue]{hyperref}
\usepackage{xcolor}
\begin{document}

\title{Simulating Curved Lipid Membranes Using Anchored Frozen Patches}

\author{James F. Tallman}

 \author{Antonia Statt}
  \affiliation{Department of Materials Science and Engineering, Grainger College of Engineering, University of Illinois Urbana-Champaign, Urbana, Illinois 61801, USA. statt@illinois.edu}

\begin{abstract}
Lipid bilayers often form high-curvature configurations due to self-assembly conditions or certain biological processes. 
However, particle-based simulations of lipid membranes are predominantly of flat lipid membranes because planar membranes are easily connected over periodic boundary conditions. 
To simulate a curved lipid membrane, one can simulate an entire vesicle, a cylinder, or a bicelle (disk-like bilayer aggregate). One can also use artificial methods to control curvature, such as applying virtual walls of beads, radial harmonic potentials, or  ``tape up the edges''.
These existing methods have limitations due to the method by which curvature is imposed. 
Herein, we propose an alternative method of introducing arbitrary curvature by anchoring a curved lipid membrane with ``frozen'' equilibrated membrane patches. The method presented here is compatible with all particle-based lipid models and easily extended to many geometries. 
As an example, we simulate curved membranes with DPPC, DOPC, DLPC and DOPE lipids as parameterized by the Martini3 coarse-grained model.
This method introduces limited finite-size artifacts, prevents lipid flip-flop at membrane edges, and allows fluctuations of the free membrane center. 
We provide verification of the method on flat membranes and discussion on extracting shape and per-leaflet quantities (thickness, order parameter) from curved membranes. Curvature produces asymmetric changes in lipid leaflet properties.
Finally, we explore the coupled effect of curvature and membrane asymmetry in both number and lipid type. 
We report the resulting unique morphologies (inducing gel phase, faceting) and behaviors (thickness dependent on adjacent leaflet type) that are accessible with this method. 
\end{abstract}

\maketitle
\section{Introduction}

Lipid molecules can self-assemble into a variety of structures due to their amphiphillic nature \cite{israelachvili1975model}, with planar lipid bilayers being one of the most studied. However, many biologically and physically significant processes occur in bilayers adopting non-planar shapes. For example, bilayers form vesicles that have radii of curvature as small as 15 nm \cite{huotari2011endosome,hurley2010membrane,marrink2003molecular}, demonstrating the constraints imposed by geometry. Mitochondrial crista junctions have radii of curvature as small as 10 nm \cite{manella2006structure,golla2024curvature,boyd2017buckling}. Similarly, lipid droplets form by the budding of lens-like oil droplets to minimize the monolayer curvature \cite{Olzmann2019,chorlay2018asymmetry,thiam2016physics}, highlighting the role of curvature in maintaining membrane organization. Membrane fusion, fission, and budding further create highly localized regions of intense curvature \cite{Grafmuller2009fusion,hurley2010membrane,gruenberg2004biogenesis,marrink2003mechanism,marrink2003molecular,kolokouris2024role}. Additionally, lipid membranes often exhibit compositional or number asymmetries which either induce or are induced by membrane curvature\cite{rival2019phosphatidylserine,doktorova2020structural}. These asymmetries and curvatures are related to functional behaviors such as membrane fusion, repair, and signaling \cite{doktorova2020structural}. These examples collectively underscore the critical role of curvature in shaping the structural and functional behavior of lipid membranes.

Fundamental understanding of lipid membranes has been developed through a combination of experimental and computational approaches. Experimental techniques provide critical insights into the structure and dynamics of membranes across nano- to mesoscopic scales, while molecular dynamics provide real-space molecular-level information into the assembly and behavior of model lipid membranes. Molecular dynamics often leverages periodic boundary conditions that constrain bilayer systems to one of three cases: a system modeling the self assembly of entire vesicles \cite{cooke2006coupling} or cylinders \cite{Sreekumari2022}, a system modeling bicelles \cite{mahmood2019curvature,mandal2020molecular,koshiyama2019bicelle, cooke2006coupling,Pohnl2024bicelle}, or a system modeling planar membranes connected over the periodic boundary condition. As discussed in the work by Foley and Deserno \cite{Deserno2024asymmetry}, membranes simulated over periodic boundary conditions cannot relax differential area strain by bending in the out-of-plane dimension, as any curvature must induce an opposite, energetically unfavorable, curvature elsewhere on the membrane.

Nevertheless, it would be advantageous to be able to simulate curved membrane systems without needing to simulate the whole vesicle (which is computationally expensive) and without simulating bicelles (which allow frequent flip-flop over the free ends of the bicelle). Recent methods to solve this problem have been summarized by Larsen \cite{larsen2022molecular}. These methods impose artificial beads or forces to break the periodic symmetry of the simulation, allowing the study of curved membranes. One approach involves adding a wall of virtual beads to enclose the membrane into a given curvature \cite{yesylevskyy2017influence,yesylevskyy2019curvature,boyd2018bumpy, golla2024curvature}. This method was used to demonstrate an increase in permeability in highly curved membranes \cite{yesylevskyy2019curvature}, allowed arbitrary geometry \cite{boyd2018bumpy}, and introduced minimal defects \cite{yesylevskyy2017influence}. Another method leverages harmonic radial constraints which enables simulation of membrane fluctuation and unimpaired lateral diffusion \cite{yesylevskyy2021encurv}. Attractive porous surfaces placed below the membrane have allowed the study of protein localization relative to curved membrane surfaces \cite{Belessiotis-Richards2019pore}.  However, in these methods, the whole membrane is affected by the curvature imposing method. Because of this, certain related behaviors such as fusion, budding, and gel transitions become difficult or impossible to decouple from the imposition of curvature.

In addition to artificial forces, there exist methods that apply artificial constraints through the boundary conditions. Membrane buckling, induced by reducing the cross-sectional area of the membrane artificially, is used to understand the effect of curvature on lipid properties \cite{domanska2024exploring,boyd2017buckling,eid2020calculating} and a mechanism behind protein curvature sensing \cite{stroh2021quantifying}. Additionally, a clever method of adding screw and glide plane symmetry along the periodic boundary condition allows the simulation of effectively curved membranes \cite{Dolan2002screw}.

A notable recent contribution to studying membrane curvature is the ``Sticky Tape'' method \cite{Deserno2024asymmetry}. Foley and Deserno offered a simple solution, ``tape up the edges'' \cite{Deserno2024asymmetry}. With a free-floating ``tape'' which prevented lipid flip-flop at the edges, their systems, which contained either compositional or number asymmetry, could relax its curvature to the lowest energy state \cite{Deserno2024asymmetry}. This method allows curvature to relax given the asymmetry present in the leaflets, and has relatively few artifacts introduced by the taped edges.  However, the method is not (to our knowledge) constructed to model systems where curvature is not relaxed to equilibrium. Additionally, this method uses the ultra-coarse-grained Cooke-Deserno lipid model \cite{Cooke2005tunable}, where lipid and ``sticky tape'' attractiveness is specific to the leaflet label, if one wanted to implement the protocol directly. The choice of lipid model in this work is justified by ``facilitating novel simulation conditions, not to serve as a template for an experimentally realizable molecular system''\cite{Deserno2024asymmetry}. The present work is, in essence, an attempt to extend and translate this model to experimentally realizable systems.

\begin{figure}
    \centering
    \includegraphics[width=1\linewidth]{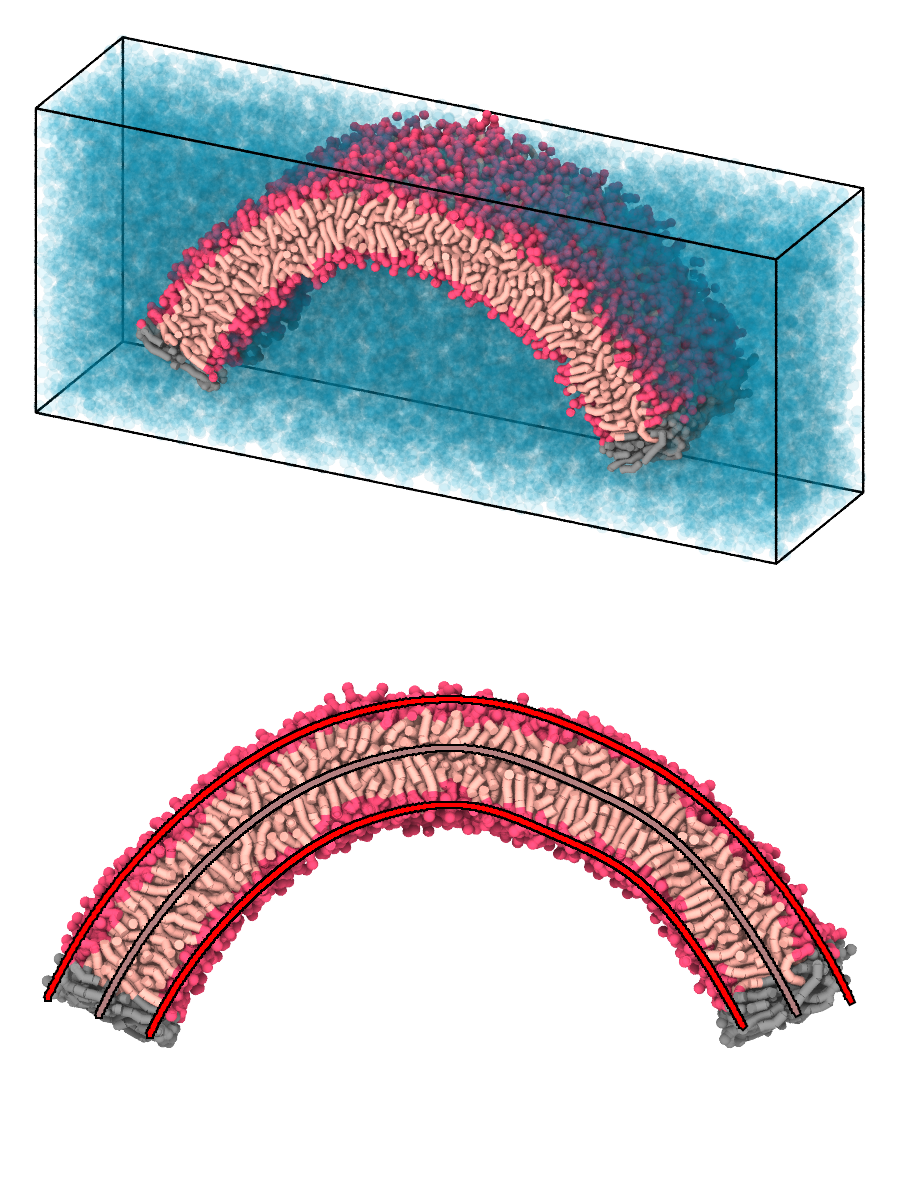}
    \caption{Snapshots of equilibrated curved lipid (DPPC) membrane with an imposed angle of $67.5^{\circ}$. Dark pink indicates lipid heads and muted pink indicates lipid tails. Grey lipids are frozen to enforce curvature. (Top) View of membrane section with explicit solvent rendered as transparent blue particles. (Bottom) View of a slice of the $xz$ plane in which solvent particles have been deleted for clarity. Outer pink lines represent curves fit to the outer and inner leaflets. Inner muted pink line identifies the membrane midplane (see Appendix \ref{app:curve-fit}). }
    \label{fig:thumbtacks}
\end{figure}

The elegant ``sticky tape'' method \cite{Deserno2024asymmetry} inspired us to develop an alternate method, which we refer to as ``frozen membrane anchors'' in this work. Those frozen membrane anchors impose a net curvature in a system, but do not define the membrane shape the bilayer must take to conform to the defined curvature. The frozen anchors of this method refer to ``frozen" patches of equilibrated lipid membrane that are not integrated forward in time. These frozen patches serve three purposes: impose the curvature, prevent flip-flop, and minimize defects.  A snapshot of the simulation equilibrated with this method is shown in Fig. \ref{fig:thumbtacks}. In this figure, the gray regions depict the frozen anchors, and the pink lipids are simulated normally.

In this work, we introduce the frozen membrane anchor method which enables the simulation of highly curved lipid membranes, agnostic to both the particle-based lipid model used \textit{and} the geometry of curvature. Through simulating flat membranes, we discuss the extent to which anchored frozen membrane patches induce artifacts. We demonstrate methods of extracting the shape of curved membranes, where are able to obtain insightful trends in the change in membrane thickness with curvature, as well as the membranes fluctuations. We decouple the leaflets and analyze how the individual leaflet properties change in response to imposed curvature. Finally, we investigate the coupled effects of membrane asymmetry and membrane curvature and the resulting unique morphologies we can access with this method. 

\section{Methods}

\subsection{Simulation Methods}
Coarse-grained molecular dynamics simulations are performed using HOOMD-blue \cite{Anderson2020} and parameterized with the Martini v3 force field \cite{Souza2021}. While this work utilizes the Martini force field, the model for imposing curvature with frozen, equilibrated membrane patches is generic to the particle-based lipid model used.

In the frozen patch method, simulations were conducted in an $NP_zT$ ensemble where the pressure in the $z$ direction was maintained at $1$ atm and box size was fixed in the $xy$ plane. As seen in Fig. \ref{fig:thumbtacks}, membranes are periodic in the $y$ dimension, curved in the $xz$ plane, where the $z$ dimension is the independent dimension which can be used to regulate the pressure of water. The temperature was maintained at $298$ K. A constant pressure barostat \cite{Martyna1994} and Nose-Hoover thermostat were used to maintain the pressure and temperature. A timestep of $20$fs was used, consistent with standard Martini simulations \cite{Souza2021}. The velocity coupling  constant ($\tau$) was 1 and the pressure coupling constant ($\tau_s$) was 12 in all simulations. Electrostatics were handled with the reaction field scheme with a cutoff distance of $1.1$ nm simulations. 

To initialize and simulate curved membranes, a multi-step protocol was implemented to minimize the equilibration time and maximize the flexibility of the model. Initially, small membrane segments, or patches, were simulated for a sufficiently long time to template equilibrium structure. These patches are periodic and flat. These patches are stitched together with imposed curvature to simulate a larger curved membrane. In these simulations, 4 patches are stitched together together, however this number can change based on the desired membrane shapes. An example workflow containing patch and curved membrane initialization and equilibration is documented here: \url{https://github.com/stattlab/bent_membranes}. This workflow relies on python package which implements the Martini v3 force field \cite{Souza2021} for HOOMD-blue \cite{Anderson2020}, \url{https://github.com/stattlab/martini3}. More specifically, the protocol follows the following steps:
\begin{itemize}
    \item \textbf{Create equilibrated membrane patches} which are used to template frozen region and curved shapes.
    \begin{itemize}
        \item Initialize $N/4$ lipids randomly in a membrane-like configuration. 
        \item Fill the rest of the box with water beads. \item Equilibrate the patch using an $NP_{xz}T$ ensemble with a pressure of 1 atm in the $z$ direction and 0 atm in the $x$ direction. The $y$ dimension is fixed. Simulate the patch for 100,000 timesteps with a timestep of 0.0002 $ps$ to mitigate initial high energy configurations. 
        \item Simulate the patch with a timestep of 0.02$ps$ for 500,000 timesteps. Following these steps, an equilibrium flat membrane configuration is created that can be used for the ``frozen" regions.
    \end{itemize}
    \item \textbf{Stitch the membrane patches together} into one curved membrane. Unwrap the lipids which cross the periodic boundary condition and place them on one side of the membrane. Additionally, tag beads on the exterior patches that are within 1.5 nm of the boundary as ``frozen" beads. When placing patches, rotate the patches in the $xz$ plane according to the angle imposed and the relative position in the cell. Translate the position in the $xz$ plane based on the defined geometry so the patch is commensurate with the other placed patches. Fill the remainder of the cell with water beads at approximately the density of Martini water. This step can be easily adapted to various different geometries. 
    \item \textbf{Minimize energy and remove overlaps} of configurations using the Fast Inertial Relaxation Engine (FIRE) algorithm \cite{Bitzek2006fire}: Perform short (100,000 timesteps, dt = 0.001) FIRE energy minimization with a displacement capped integrator, purely repulsive pair interactions, and harmonic bond/angle potentials, integrating forward in time all non-frozen lipids.
    \item \textbf{Equilibrate the curved membrane} in an $NP_zT$ ensemble and 1 atm of pressure in the z dimension, applying the Martini force field (Lennard-Jones (12-6), bond harmonic, cosine angle harmonic, and coulomb forces via reaction field) forces with a timestep of 0.002 ps for 10,000 timesteps. Integrate forward in time all non-frozen beads
    \item \textbf{Simulate the curved membrane} in an $NP_zT$ ensemble and 1 atm of pressure in the z dimension, applying the Martini force field with a timestep of 0.02 ps for 1,000,000 timesteps. Membrane thickness equilibrates after roughly 20,000 timesteps. Integrate forward in time all non-frozen beads. 
\end{itemize}

Central to this method is the pre-equilibration of the membrane morphology \textit{prior} to curved membrane simulation. Other details, such as using membrane patches for said equilibration (as opposed to projecting a single equilibrated large membrane onto a cylinder), force field of choice, or the exact lengths of equilibration and simulation, are not central to the method, they are simply one option that ensured equilibrated results for the lipids and properties investigated here.  

\subsection{Simulated Lipids}

In this work, four different phospholipid species with unique intrinsic curvature \cite{Kaltenegger2021intrinsic} were investigated: 1,2-dilauroyl-sn-glycero-3-phosphocholine (DLPC, 12:0 PC), 1,2-dipalmitoyl-sn-glycero-3-
phosphatidylcholine (DPPC, 16:0 PC), 1,2-dioleoyl-sn-glycero-3-
phosphatidylcholine (DOPC, 18:1 PC), and 1,2-dioleoyl-sn-glycero-3-phosphoethanolamine (DOPE, 18:1 PE), listed in order from most positive to most negative intrinsic curvature. Experimentally, all these lipids show similar apparent bending moduli when in the fluid phase\cite{Nagle2017moduli}.  These lipids are parameterized with the commonly used Martini v3 force field\cite{Souza2021}. 
In Martini simulations, DPPC is the lipid closest to its gel phase temperature, contributing to its distinct behavior \cite{wang2016dppc,sharma2021evaluating}. DLPC, on the other hand, has a shorter hydrocarbon chain with fewer carbon atoms compared to the other lipid species, affecting its packing properties and further supressing the gel phase \cite{silvius1979thermotropic}. The membranes formed by DOPC and DOPE are less rigid due to the presence of unsaturated hydrocarbon tails, which introduce kinks and disrupt tight lipid packing \cite{eid2020calculating}. 
Simulating four lipid types with unique properties demonstrates the versatility of the method and allows us to explore different regimes of curved membranes.

\subsection{Characterization of Flat Membranes}

To calculate total membrane thickness, spanning the entire membrane \cite{nagle2000structure}, the $z$ position of the lipid head bead of the phospholipids on the upper and inner leaflets were collected and averaged over an area, then subtracted from each other. For average thicknesses, this area was the whole membrane. For thickness maps, the grid is constructed such that all pixels have an area of $1 \text{nm}^2$.

To quantify the average alignment for every lipid in the simulation, the sn1 and sn2 order parameters were found by taking the angle $\theta$ between the vector defined by the second and third hydrophobic beads in each respective lipid tail and the membrane normal. The membrane normal was assumed to be the [001] direction. The order parameter was calculated using $S = \left(3\langle\cos^2{\left(\theta\right)}\rangle - 1\right)/2$. The maps are created from calculating an average value with a pixel size of $1nm^2$ and 60 frames (where a frame is written every $0.2$ ns).

\subsection{Analysis of Curved Membranes}

Two assumptions of flat membranes fail when moving to curved membranes, rendering the above methods difficult to use: it is more challenging to assign lipids to leaflets due to curvature, and membranes are not symmetric. Because the mid-sections of the membranes are freely fluctuating, the membranes do not even need to be radially symmetric in the $xz$ plane despite that geometry being imposed (by number of lipids and angle at the boundary). Here, two methods to analyze curved membranes are investigated. First, a surface mesh is used to calculate average membrane thickness and average radius. Second, the periodicity in the $y$ direction is leveraged to fit the data to a single function which models both the imposed curvature and the undulations on the membrane segments.

\subsubsection{Average Membrane Thickness}

In this method, the curved leaflet is modeled as, on average, a partial cylindrical segment with a finite thickness. The radius and thickness of this geometric object can be calculated from its surface area and volume. This relationship is shown in Appendix \ref{app:geometric}.

Using the visualization and analysis software OVITO \cite{Stukowski2009}, a surface mesh can be constructed which determines the area and volume with only one input parameter, the probe sphere radius. The average thickness can be calculated directly using this area and volume. A probe sphere radius of 0.7 nm matches the membrane thickness of flat membranes calculated from distance between head beads, and thus is used.

\subsubsection{Curve-fitting Curved Membranes}
First, all lipid beads are projected into the $xz$ plane yielding an array of two-dimensional data that represents the curved membrane state. In this study, three sets of data points are fitted: the outer leaflet, inner leaflet, and the membrane midplane. The midplane data points are obtained by taking the last (tail) bead of each lipid in the simulation as data points. The inner and outer leaflet populations are obtained using each lipids first (head) bead. To identify which leaflet a given head belongs to, the set of all lipid head beads are sorted into two populations using the Density-Based Spacial Clustering of Applications with Noise (DBSCAN) algorithm \cite{DBScan} (with epsilon = 0.8, minsamples = 2). This clustering algorithm failed to accurately separate clusters with distinct densities in the case of number asymmetric leaflets, as it assumes a constant density within clusters. For this reason, the Hierarchical DBSCAN (HDBSCAN) \cite{campello2013density}, which is developed specifically to find clusters of varying densities, was used for asymmetric membranes. 

An interpretable equation for the shape which the data should obey is
\begin{equation}
z_{tot}(x) = z_{circ}(x) + z_{cos}(x) + z_{sin}(x) + z_{adj}(x) \quad .
\end{equation}
Each of these terms is explained in detail in Appendix \ref{app:curve-fit}.
This method is similar to other methods used in the literature\cite{Nickahorn2024curv}, but follows a different functional form to account for the spherical shape and boundary conditions.

Fitting the midplane and the outer and inner leaflet yields curves which define the shape of the membrane. The membrane thickness and leaflet thicknesses were defined as the average minimum distance between two of the curves. Only the inner $3/5$ths of the curves were used for thickness calculation to avoid capturing the effects of the frozen beads.

\subsubsection{Determining Local Order Parameter}
For flat membranes, the order parameter is calculated by assuming the membrane normal is in the [001] direction. To measure the order in highly curved systems, a ``local order parameter" quantity was used which contains similar information to the lipid order parameter for flat membranes. To calculate this quantity, the lipids of each leaflet are separated (using DBSCAN or HDBSCAN) and for each lipid, an orientation vector is calculated, defined as the average vector between the tail beads and the head bead. 

For each lipid $i$, all $N$ neighboring  lipids within $0.5$ nm are identified. For each neighbor $j$, the angle $\theta_{ij}$ between itself and the central lipid $i$ is computed. The per-lipid order parameter is then calculated using the nematic order formulation 
\begin{equation}
    S_i = \frac{3\left(\frac{1}{N} \sum_{j=1}^N\cos^2\theta_{ij} \right) - 1} {2} \quad .
\end{equation} This quantity $S_i$ is then averaged over all lipids in a leaflet to give a system-wide average ``local order parameter''.

\subsection{Extracting fluctuations and apparent bending modulus}

The curve which defines the membrane shape consists of two components: an average shape and fluctuations. These fluctuations are added onto the arc of the membrane with the imposed periodic boundary conditions. The second moment of these fluctuations can be calculated and used to approximate the bending modulus \cite{deserno2007fluid, chaurasia2018evaluation}.
This relationship describes the fluctuation amplitudes, $h$, in terms of the wave vectors, $q$, and the apparent bending modulus $\kappa$,
\begin{equation}
    \label{eq:fluct}
     \langle h^2 \rangle = \frac{k_BT}{4\pi \kappa} \frac{q_{max}^2 - q_{min}^2}{\left(q_{min}q_{max}\right)^2} \approx \frac{k_BT}{16 \pi^3 \kappa}L^2 \quad .
\end{equation}

This apparent bending modulus measurement captures the ability for membranes to deform away from their equilibrium shape by undergoing bending behavior. However, it originates from an approximate theory and is severely effected by the finite size of the simulated system. For those reasons, it should be interpreted as an apparent bending modulus and not an exact value.

\subsection{Creating membranes with compositional and number asymmetries}

The creation of membranes with composition or number asymmetry was done entirely in the first step of membrane creation: patch initialization. Instead of creating a leaflet with equal number or types of lipid on the outer and inner leaflet, mismatch was created and equilibrated. For number asymmetry simulations, the number asymmetry is defined as $({n_o - n_i})/({n_o + n_i})$ where $n_i$ is the number of lipids on the inner leaflet and $n_o$ is the number of lipids on the outer leaflet.

In SI Figs. 8-11, it can be seen that at high asymmetries, several membranes failed in this patch equilibration phase leading to malformed frozen patches and membranes.  Lipid flipflop did not occur except for those cases of membrane failure (pore formation, buckling, rupture). To confirm the stability of morphologies and improve confidence on the absence of lipid flip-flop coupled number asymmetry and membrane curvature simulations were ran for a total of 0.2 $\mu s$. 

\section{Results and Discussion}

\subsection{Frozen Patches Do Not Introduce Significant Membrane Defects}
\begin{figure}
    \centering
    \includegraphics[width=1\linewidth]{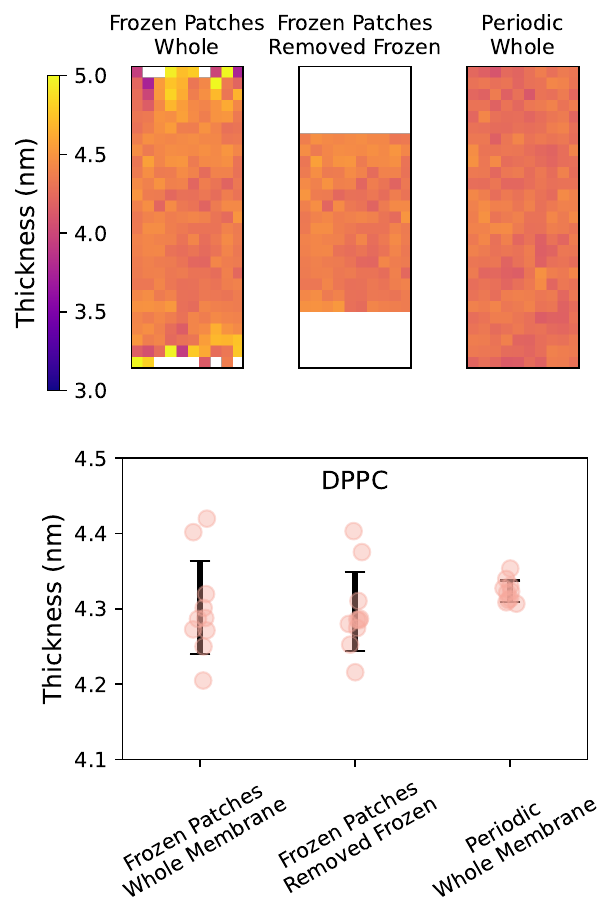}
    \caption{Effect of frozen membrane anchors on equilibrium thickness of DPPC membranes. (Top) Average thickness of frozen anchored DPPC membranes, frozen anchored membrane only counting the ``middle" region, and periodic membrane, all averaged over 60 frames. (Bottom) Average thickness of ten independently initialized simulations, averaged over 60 frames.}
    \label{fig:finite-size-thickness}
\end{figure}

In this section, we investigate how the presence of frozen patches influences the local and average membrane properties.
The method introduces several artifacts by the nature of the frozen patches being placed in the simulation. 
The dynamics of lipids near frozen patches are slower than those in the bulk membrane, as shown in SI Fig. 4. 
Similarly, frozen anchored patches will lead to different allowed fluctuation modes compared to those which exist on a periodic, cylindrical, or spherical shape. 
Aside from the membrane fluctuations, these anchored patches primarily affect the dynamics and structure of only their local environment. 
The lipids in the middle of the membrane do not ``feel'' that they exist in a partial membrane section, allowing accurate sampling of high-curvature membrane behavior.

To show the extent to which freezing the membrane ends alters the properties of the membranes in equilibrium, we designed control simulations. In these simulations, membrane patches were equilibrated and then placed them either into a simulation with periodic boundary conditions or into a non-periodic, anchored membrane. The initial membrane configuration is identical between the two methods. These membranes were then simulated for $0.02\mu s$ (1,000,000 timesteps) and the equilibrated thickness and order parameter were measured. This procedure was repeated 10 times from independent starting configurations. In addition to the data presented, simulations of patches of different sizes and equilibration times were tested and displayed similar results.

In Fig. \ref{fig:finite-size-thickness} (bottom), the average thickness for each of the 10 randomly initialized membranes is shown. It is apparent that the anchored membranes sampled a broader distribution as compared to the periodic membranes; however, they exhibited similar average thicknesses. This same analysis was completed for DOPC, as shown in SI Fig. 1, and the same conclusions can be drawn. When comparing the local average thickness of one membrane with frozen patches to the periodic variant, the anchored membrane (Fig. \ref{fig:finite-size-thickness} upper left) clearly has artifacts at the frozen boundary and potentially about 1-2 nm into the membrane. These artifacts do not appear in the periodic membrane (Fig. \ref{fig:finite-size-thickness} upper right). 

Nonetheless, the artifacts in the frozen anchored membranes resembled an instantaneous thickness. This is because those regions are not integrated forward in time and thus preserve the state they were in prior to being frozen. Membranes exist with fluctuating compression and extension modes. If that membrane was in a compressed or extended state when the edge patches were frozen, the entire membrane is now constrained to that small compression or extension. This is an inevitable (albeit not that costly) effect of keeping the membrane area exactly constant. Sampling sufficient replicates will yield the correct average behavior with slightly higher variance (Fig. \ref{fig:finite-size-thickness}). This effect also highlights the importance of properly equilibrating the membrane patches prior to freezing, as well as excluding them from the analysis. In curved membranes, those small compression/tension effects can be alleviated by changing the net curvature. The effect of constant area is the largest in flat and relatively small membranes, which are the conditions tested and shown here. Additionally, there is no difference between the thickness of the center of membrane as compared to the frozen patches on average. The same trends are observed for the order parameter (shown in the SI Figs. 2 and 3), where there is no average difference between the ordering of the frozen anchored membrane, inner section of frozen anchored membranes, and the periodic membrane.

\subsection{Extracting Curved Membrane Shape}

To understand the effects of curvature, we require methods to analyze the shape and size of the membrane, which exists on a non-planar manifold. This renders many of the traditional analysis methods (thickness, density) difficult to apply. In previous publications, different approaches have been taken: distance from imposed membrane midline \cite{yesylevskyy2021encurv}, assuming spherical shape \cite{Deserno2024asymmetry} and approximate surface methods \cite{Bhatia2019Mem}. 

Here, two methods to assess membrane shape are compared. First, using the assumption of a cylindrical surface (which is the imposed geometry calculated from number of lipids and location of pinning points), thickness of the membranes was computed as a function of effective cylinder area and volume. Second, the lipid heads were treated as two dimensional data points $(x,z)$ and fitted to a curve defining the membrane surface $z(x)$. Average quantities can be calculated by operating on the curves.

\begin{figure}
    \centering
    \includegraphics[width=1\linewidth]{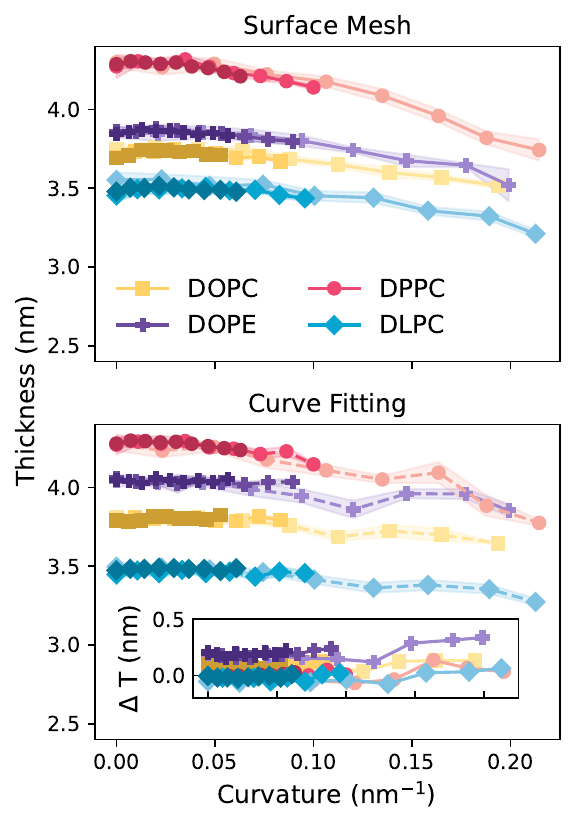}
    \caption{Membrane thickness measured using the mesh method (top) and the curve fitting method (bottom) as a function of imposed radius of curvature for DPPC  (red circles), DOPE (purple pluses), DOPC (yellow squares) and DLPC (blue diamonds) lipid membranes. The three color shades represent small (light color), medium, and large (darker color) membranes consisting of $\sim $600, 1200, and 1800 lipids respectively. A transparent region is drawn around the data to indicate the standard deviation of these quantities over 5 uniquely initialized membranes. Inset: Difference in thickness between curve fitting and surface mesh method.}
    \label{fig:curv-thickness}
\end{figure}

In the top panel of Fig. \ref{fig:curv-thickness}, the membrane thickness is plotted as a function of the inverse radius of curvature ($1/r$), measured via the surface mesh method. The radius of curvature allows the comparison across different system sizes. All membranes show similar behavior as a function of radius of curvature regardless of the simulation size. In this plot, the expected thinning of the membrane at high radius of curvature was observed \cite{yesylevskyy2021encurv,domanska2024exploring}. Also, a small thickening of the membranes when introducing a small amount of curvature is seen. This measured thickness relies on the assumption of cylindrical geometry. However, the membranes were only frozen at the edges and were otherwise freely fluctuating, meaning that while a cylinder is a realistic average geometric model of membrane shape, it is not an exact model. As shown in SI Fig. 5, membranes did adopt asymmetric shapes which severely violate this assumption. Additionally, this method can easily be used to model non-cylindrical geometries which will not follow this geometric construction. Thus, while the geometric method is a relatively accurate first-pass at the membrane shape, a more generic method to approximate the membrane shape was also developed here.

\begin{figure}
    \centering
    \includegraphics[width=1\linewidth]{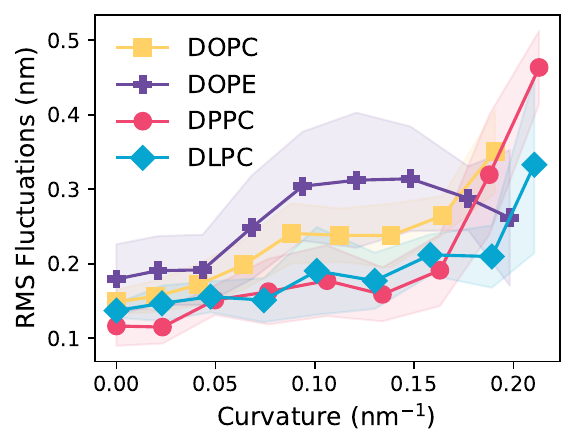}
    \caption{Measure of membrane fluctuation as a function of imposed curvature. Membrane fluctuations are reported as the second moment of the fluctuations which exist on the curved membrane structure. Membranes are reported with the smallest box size ($N \sim 600$), values are averaged over 80 frames and error are represented as the standard deviation of five uniquely initialized snapshots.}
    \label{fig:fluctuations}
\end{figure}

In the bottom panel of Fig. \ref{fig:curv-thickness}, the thickness of the lipid membranes as measured by the curve fitting method is shown. Again, the expected thinning of membranes at high curvature as well as a relatively unchanging thickness at low curvature was observed.  
Interestingly, the slight thickening at low curvature observed with the surface mesh method disappeared with this mode of analysis. 
This implies that these curved membranes do not exist as exact cylinders and consequently they should not be treated as such despite the fact that the cylinder is the ``imposed" curvature. 
Membrane thicknesses were similarly ordered between lipid types but the absolute values change for DOPE where the head groups are pulled further into the solvent, a detail strongly detected by the curve-fitting which analyzes only the head beads and less-so by the mesh which constructs an average surface. This difference is highlighted in the inset figure.
Additionally, the lack of change in membrane thickness at low membrane curvature suggests a ``linear-response'' regime, where the outer leaflet thins and the inner leaflet thickens by equal but opposite amounts. That ``linear-response'' regime ends at relatively low curvature (radius of $\sim$ 40 $nm$).
All lipid bilayers with no curvature share qualitative agreement with experimental data \cite{Kucerka2011fluid} and are representative of the Martini model \cite{Souza2021}.

\begin{figure}
    \centering
    \includegraphics[width=1\linewidth]{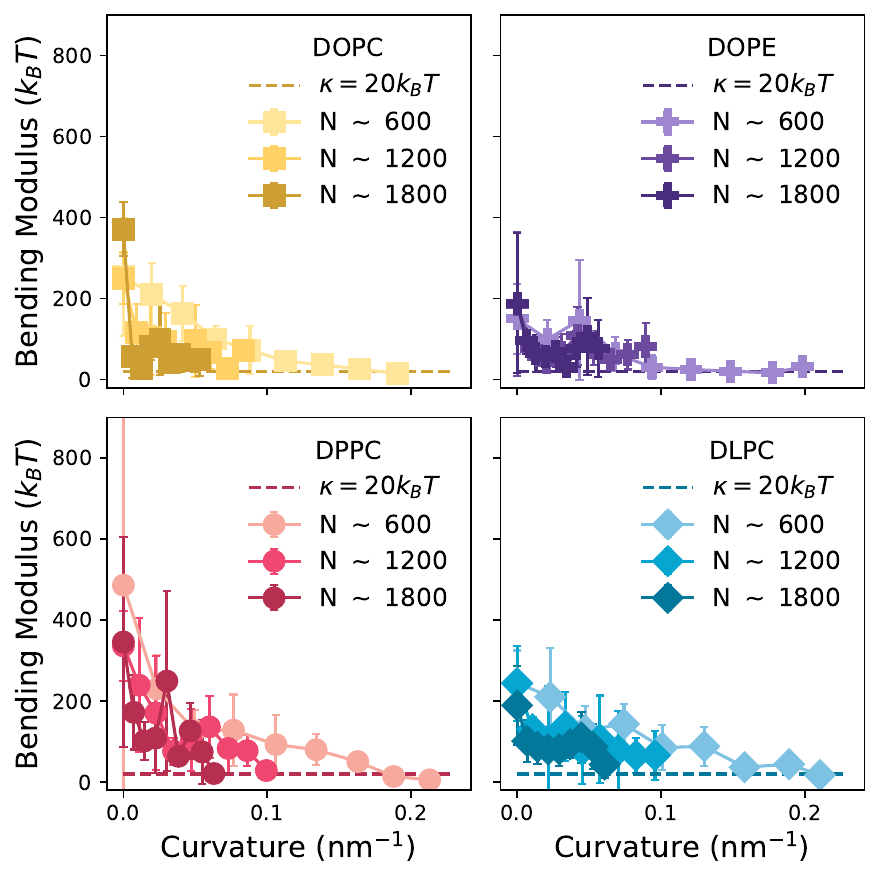}
    \caption{Apparent bending modulus of lipid membranes as a function of curvature. The four panels show the dependence on curvature of each lipid species. From lightest to darkest, three differently sized systems are reported ($N \sim 600$, $N \sim 600$, $N \sim 1800$). Apparent bending modulus is obtained from an approximate relationship with bending modulus and membrane size. Values are averages of five independently initialized membranes and error bars represent the standard deviation of those values.}
    \label{fig:fluctuations_modulii}
\end{figure}

With an appropriate way of assessing the non-cylindrical membrane shape, properties such as membrane fluctuations can be extracted. Once the underlying circle which these membranes live on is subtracted, the remaining curve is a spectra of the fluctuation on the membrane surface. Examples of these fluctuations are shown in SI Fig. 6, where it is apparent that membranes fluctuate more in the membrane center than at membrane edges. Taking the root mean squared value of the total fluctuations over the membrane arclength, shifted such that the first moment is zero, yields a measure of the rigidity of the membrane. From the Helfrich model, the second moment of these fluctuations can be related directly to an ``apparent'' bending modulus \cite{deserno2007fluid} (see eq \ref{eq:fluct}). 

In Fig. \ref{fig:fluctuations}, the root mean squared fluctuations of membranes for the smallest membrane size are shown. All membrane sizes are shown in SI Fig. 7. The ordering of the fluctuation amplitudes and apparent bending modulus is as expected, with saturated lipids fluctuating less than unsaturated lipids. DPPC is stiffer than DLPC due to having longer chains. As membrane curvature increases, the membranes decrease their ``rigidity" and fluctuate more. One possible explanation for this can come from the trivial example of holding a sheet of paper between one's fingers. If the paper is flat, the paper is not able to deform easily. However, if the sheet of paper is held at an angle on both sides (akin to how the membranes are curved in this work), the center can easily be disturbed in the direction perpendicular to the pinned edges. Thus, we speculate in these finitely sized systems, bending can activate suppressed fluctuation modes.

From a first order theory of the bending moduli,  bending moduli of membranes can be extracted from the amplitude of the fluctuations. This analysis is shown in Fig. \ref{fig:fluctuations_modulii}. In this figure, for each lipid species, the bending moduli decreased as the imposed curvature is increased, consistent with our previous explanation. Likewise, as the membrane size was increased, there was also a decrease in bending moduli. This is a similar way of ``activating" certain longer-wavelength bending modes which are artificially frozen-out by pinned membrane edges (or periodic boundary conditions). Additionally, these values appear to plateau at high curvatures. 

Experimentally, bending modulus can be obtained using different characterization methods which probe membranes at different length and time scales \cite{Gupta2019dynamics}. Neutron spin echo (NSE) spectroscopy \cite{chakraborty2020cholesterol}, flicker spectroscopy \cite{fricke1986flicker,chaurasia2018evaluation,elani2015measurements}, diffuse X-ray scattering \cite{lyatskaya2000method}, micropipette aspiration \cite{henriksen2004measurement}, and electric deformation \cite{gracia2010effect} are commonly used. Nagle \cite{Nagle2017moduli} reports bending moduli of three of the four lipid species studied in this work, measured with diffuse X-ray scattering of lipid stacks at the micron scale. For example, in the fluid phase, DPPC (at $50^\circ C$), DOPC (at $30^\circ C$) and DLPC (at $30^\circ C$) have a bending modulus of $18 k_BT$, $16.3 k_BT$ and $20.4 k_BT$ respectively.\cite{Nagle2017moduli}. The dashed lines in Fig. \ref{fig:fluctuations_modulii} indicate the experimentally relevant scale of bending modulus of about $\sim 20 k_BT$. It is interesting that these apparent bending moduli measurements obtained from the membrane fluctuation plateau around the experimentally measured values.

\subsection{Individual Leaflet Response to Curvature}

The lipid membrane response to curvature is the combination of the deformation of the bottom and top leaflets. These two leaflets would be expected to behave uniquely as the inner leaflet is in a state of compression while the outer leaflet is in tension. With the method to fit the shape of the leaflets and the membrane midplane as presented here, the thickness and ordering of the two leaflets can be extracted independently.
The deformations of these leaflets can be related to the previous hypothesis that there are two distinct deformation regimes in curved lipid membranes.

\begin{figure}
    \centering
    \includegraphics[width=1\linewidth]{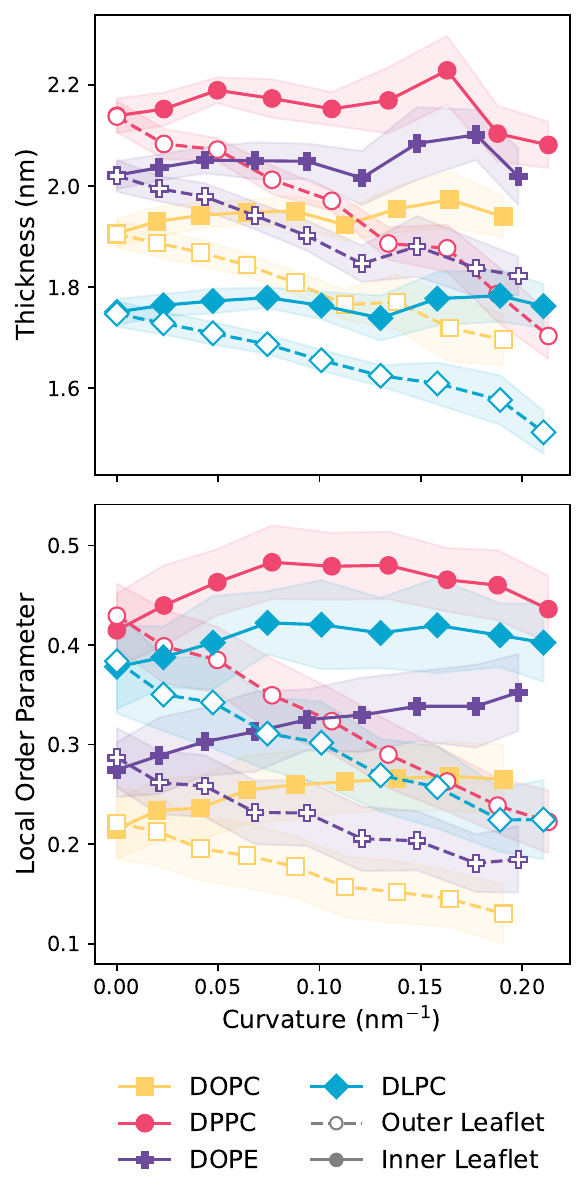}
    \caption{Effect of curvature on leaflet thickness (top) and local lipid order (bottom) for DOPC (yellow squares), DPPC (pink circles), DOPE (purple plusses), and DLPC (blue diamonds). Open circles and dashed lines indicate the outer leaflet. Closed circles and solid lines indicate the inner leaflet. Averages and standard deviations were obtained from 5 uniquely initialized simulation. Leaflet thickness was obtained with the curve fitting method. Order parameter was obtained with local order measurement. Simulations were conducted with $\sim$ 600 lipid molecules.}
    \label{fig:thickness_order}
\end{figure}

In Fig. \ref{fig:thickness_order}, the changes in the individual leaflet thickness and order are reported as a function of radius of curvature for the four lipid types. Examining the thickness of the individual leaflets, a physical explanation for the `linear' regime becomes apparent: the outer leaflet thickens at the same rate the inner leaflet thins. When the membrane is only slightly deformed, this follows from the equivalence of the stretching modulus and the compression modulus near equilibrium. 

However, it is clear that the outer leaflet deviates from linear behavior more prominently than the inner leaflet at larger curvatures. Decreasing the thickness of a leaflet implies lipids, on average, need to be more disordered. This provides a rationale for why DPPC (and also DLPC) appear to display a greater change in overall thickness as compared to the unsaturated lipids. The membrane thinning of the outer leaflet for all lipid types appears to increase in magnitude as curvature increases. Physically, this behavior could correspond to the approach to a minimal thickness before rupture. As the saturated lipids are highly ordered in equilibrium, the compression quickly reaches a point of maximum order and thickness given the applied curvature.

In addition to analyzing the leaflet thickness, the average local order of each leaflet was obtained by comparing each lipids orientation to all of its neighbors for every lipid in the simulation. DPPC is the most ordered, with DLPC and DOPE being less ordered, and DOPC being the least ordered. This is expected by the properties of the lipids. At zero curvature, each leaflet in the homogeneous membrane of one lipid type has similar order. When imposing small curvature, there is a linear regime with similar magnitude of slope for the inner and outer leaflet, similar to the observed changes in leaflet thickness. At higher curvature, the inner leaflet plateaus to a maximum order (except for DOPE) while the outer leaflet continues deordering. DPPC membranes exhibit an increased rate of loss of order compared to all other lipid types tested. The loss in order of the inner leaflet is nearly-perfectly linear with imposed curvature.

\begin{figure}
    \centering
    \includegraphics[width=1\linewidth]{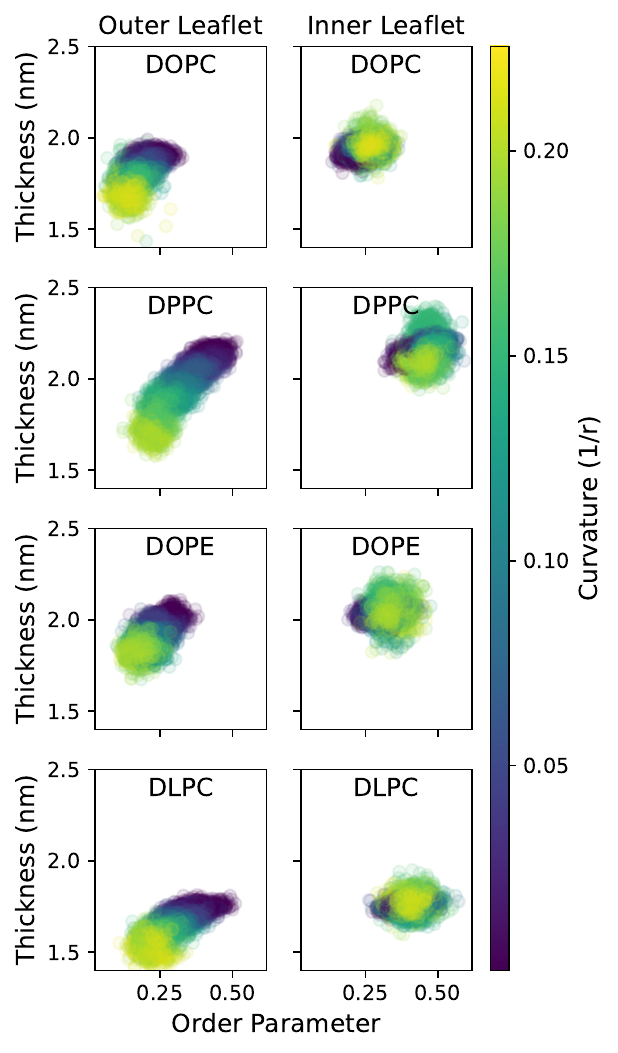}
    \caption{Correlations between thickness, curvature, and order parameter for the inner and outer leaflet of each lipid type tested. Each dot represents on average membrane quantity from an equilibrated frame. Five uniquely equilibrated trajectories were analyzed for each lipid type.}
    \label{fig:leaflet_thickness}
\end{figure}

The anchored frozen patches method allows us to impose some curvature to a system and measure an instantaneous average thickness, net curvature, and order parameter. In Fig. \ref{fig:leaflet_thickness}, these three calculated quantities are visualized for every frame of the equilibrated membranes, where the left column shows the outer leaflet behavior and the right column shows the inner leaflet behavior. It is obvious that the thickness, curvature, and order parameter are strongly correlated in the outer leaflet when deformation is imposed. DPPC exhibits the largest change from no curvature to highly curved systems, likely because it is the most ordered state initially. DOPE, which is initially the second thickest, does not change its properties nearly as much, suggesting the zero curvature ordering is more important for the observed correlations than the zero curvature thickness.

However, when analyzing the correlations between thickness and order parameter for the inner leaflet, the correlations are less apparent than for the outer leaflet. As curvature increases, the point clouds shift towards more ordered states with higher thickness. These correlations are much weaker. In DPPC, it is clear that the membrane thickens and then thins as curvature is increased.

Our analysis shows that when a membrane is curved, the inner and outer leaflets experience different stress states and deformations. The outer leaflet thins dramatically, while the inner leaflet compresses marginally. Bending a membrane creates asymmetric lipid properties.  
% This suggests that a possible reason that a cell might create an asymmetric membrane (in either composition or number asymmetry) is to expose different lipid types to different stress states and promote unique and functional behavior. 

\subsection{Simulating Membranes with Both Curvature and Number or Type Asymmetry}
\begin{figure}
    \centering
    \includegraphics[width=1\linewidth]{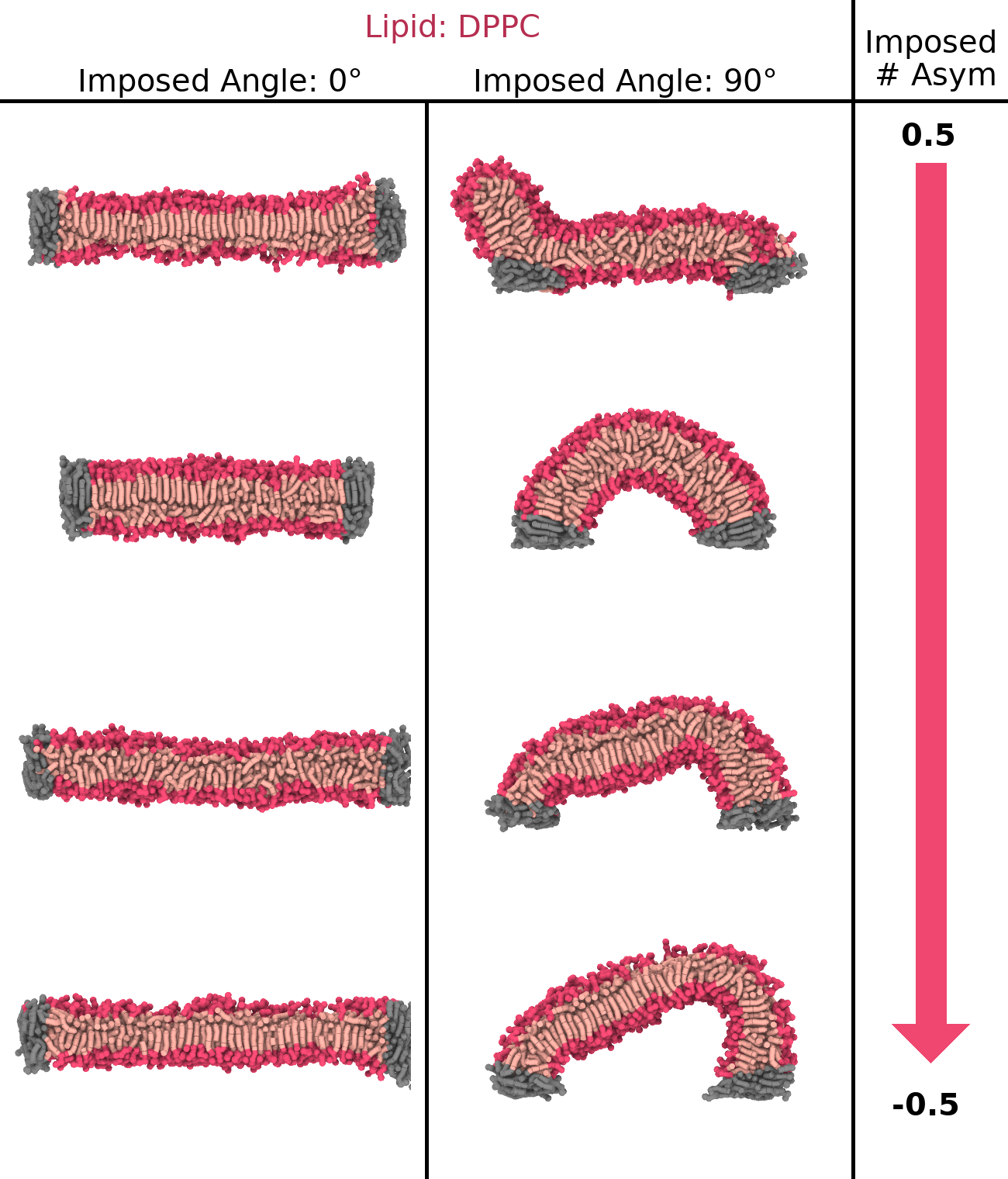}
    \caption{Renderings of DPPC membrane simulations with imposed net curvature (left column corresponds to membranes with 0$^\circ$ imposed curvature, right column  corresponds to membranes with $90^\circ$ imposed curvature) and imposed number asymmetry (top to bottom 0.5,0.2,-0.2,-0.5). Images are rendered after simulating for 0.2$\mu s$.}
    \label{fig:DPPC_number}
\end{figure}

Naturally forming membranes are asymmetric \cite{devaux1991static, lorent2020plasma}, and membrane asymmetry and curvature are fundamentally related \cite{Deserno2024asymmetry,Zhu2024}. Chaisson and coworkers identifed four main approaches to simulating asymmetric membranes, which involve controlling the number of lipids on each leaflet, the areas, the differential stress, or mimicking ``biologic asymmetry" \cite{chaisson2024asym}. By activating certain bending behavior not accessible in periodic membrane simulations, and by not controlling the membrane area directly, we speculate that the frozen patches method could provide an interesting set of new boundary conditions to investigate asymmetric membranes.

Here, the frozen patches method is used to investigate the coupled effects of curvature and membrane asymmetry. We investigate both number asymmetry and type asymmetry. This curvature-number asymmetry coupling has been shown with a DPD-type lipid model previously \cite{Sreekumari2022}. The DPD model has softer interactions and less degrees of freedom and thus flip-flop was observed on their time-scale of simulation. Likewise, Foley et al. \cite{Deserno2024asymmetry} showed the ability to equilibrate the curvature of both numerically and compositionally asymmetric membranes with their sticky tape method. With the frozen patches method, one is able to impose arbitrary curvature on highly asymmetric membranes and observe the morphological changes as a result. Because the method allows the simulation of smaller systems, this analysis can be performed readily on a less coarse-grained, more expensive Martini model, providing insights into the relative phase behavior for various lipid types. 

The protocol to create these membranes mimics that of the symmetric membranes, however, the initial patches are equilibrated for a short time with asymmetric distribution of lipids instead of a symmetric distribution. Once a membrane patch is stitched into the full curved membrane, flip-flop does not occur (as seen in SI Fig. 13) on the timescale of the simulations presented here. If the asymmetry is too high in the initial patch equilibration, pores or non-bilayer membrane structures might form and lipids might flip-flop. Hence, a number asymmetry in this section refers to the imposed number asymmetry and not the equilibrated number asymmetry. Number asymmetric membranes are simulated for ten times as long as symmetric membranes (0.02 $\mu s$) to ensure no long-term behaviors such as lipid flip-flop are present.

In Fig. \ref{fig:DPPC_number}, representative morphologies for simulated number asymmetric DPPC membranes with and without imposed net curvature are shown. The accessible morphologies are diverse, unique, and influenced by the net curvature of the system.
At zero net curvature and high number asymmetry, the more concentrated outer leaflet (on the top  of the figure) transitions to the ordered gel phase. The inner, less concentrated  leaflet forms an unstructured monolayer of lipid on the opposite side of the gel phase. At moderate asymmetry, leaflet thicknesses differ slightly, but both leaflets remain fluid.
Interestingly, with an imposed net curvature and an over-packed outer-leaflet, a protrusion appears. The protrusion consists only of lipids originating from the outer leaflet. These protrusions are more common in systems with more fluid membranes of DOPC or DOPE, as seen in SI Figs. 8-11. With moderate positive asymmetry in DPPC systems, the two leaflets seem to be structurally similar. With a dense inner leaflet, the inner leaflet also forms a gel phase, leading to faceted or angled membrane configurations.
For completeness, an example snapshot from each asymmetry, lipid type, and asymmetry is shown in the SI Figs. 8-11.

At standard conditions, the melting transition temperature ($T_m$) of an experimental DPPC bilayer is at $\sim 41^\circ C$ \cite{chen2018determination,Grazyna2021phase}. In the Martini model, however, simulations conducted at $2.47 k_BT$ or $25^\circ C$ using the martini model still yield DPPC in the fluid phase, because of the inability of a coarse-grained model to capture all details to the precision of experimental systems \cite{wang2016dppc,sharma2021evaluating}. The ability for membrane number asymmetry to cause the transition to the gel phase selectively in a single leaflet is a physical manifestation of the large internal stresses that can accompany both curvature and asymmetry. This encourages further investigation into structural changes of lipid membranes when membranes undergo rapid changes in curvature without sufficient time to relax number asymmetry.

The structural diversity of membranes with coupled asymmetry and curvature is stark. Simulations were conducted at 2.47 $k_BT$ where, in the Martini force field, minimal appearance of the gel phase is expected. Regardless, with large number asymmetries imposed, the formation of gel phases is clearly observed inside of DPPC membranes. This gel phase transition was not seen in other lipid types (see SI for DOPE, DOPC, DLPC). The shape of the DPPC membrane with faceted angles is reminiscent of structures seen in experimentally synthesized DPPC vesicles at room temperature \cite{jagalski2016biophysical} and simulated buckled gel phase membranes \cite{nagarajan2012dynamics}. The protrusion structures are similar to those shown previously \cite{Sreekumari2022}  when coupling asymmetry and curvature was investigated, and are similar to structures formed by simulating full cylinders with large number asymmetry (SI Fig. 15).

\begin{figure}
    \centering
    \includegraphics[width=1\linewidth]{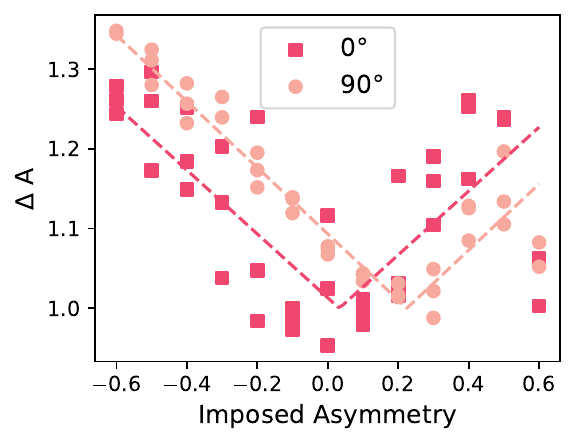}
    \caption{Comparison of DPPC membrane area (calculated via OVITO surface mesh) as a function of imposed asymmetry for flat membranes ($0^{\circ}$) and highly curved membranes ($90^{\circ}$). }
    \label{fig:DPPC-area}
\end{figure}

In Fig. \ref{fig:DPPC-area}, the average area of the surface mesh constructed around a hybrid membrane is reported against the imposed asymmetry. 
This area contains information of membrane geometry and the effect of curvature. A membrane has the minimal area when both leaflets share equal packing.
The minimum is at 0 asymmetry for an angle of $0^\circ$ and at 0.2 for an angle of $90^\circ$. 
These minimum asymmetry corresponds to a ``relaxed" membrane area where neither leaflet is causing a mismatch in tension, and is similar to the calculated asymmetry obtained to recover equal lipid density for each leaflet. 
Outliers are present at extreme imposed asymmetries where either protrusions/pores appear or the initial configuration could not sustain the asymmetry. 
It is noteworthy that DPPC membranes can maintain significant asymmetry without losing a bilayer configuration.

In SI Fig. 12, this same plot is presented for all four lipid types, although other lipid types fail to sustain as high of a number asymmetry and thus have more outlying data points.
In SI Fig. 13, the final asymmetry of membrane related to the initial asymmetry is presented. It is clear that flip-flop only occurs in membranes which do not form bilayer structures during the initial patch equilibration. The range of bilayer maintaining number asymmetries is largest for DPPC and smallest for DLPC, possibly related to the bilayer thickness.
In SI Fig. 14, the local order parameters for the outer and inner leaflets that are both curved and flat are shown and discussed. DPPC shows the strongest relationship between ordering, curvature, and number asymmetry. 
\begin{figure}
    \centering
    \includegraphics[width=1\linewidth]{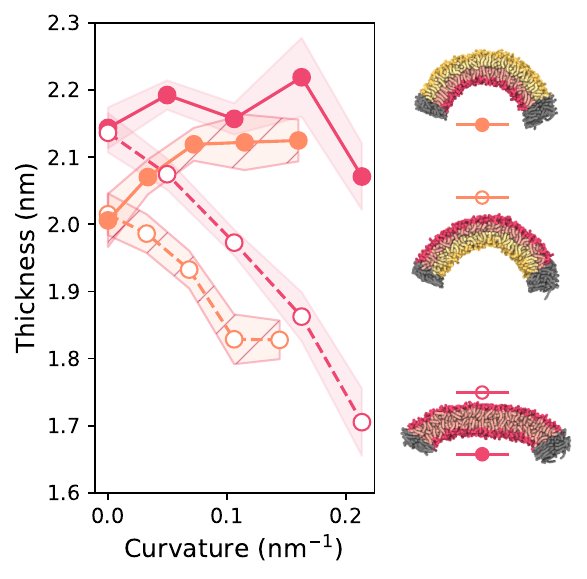}
    \caption{(Left) Thickness of DPPC leaflets (measured from head group to membrane midplane) with adjacent leaflet being composed of either DPPC (pink) or DOPC (orange).  Data presented is average of three randomly initialized replicates for each lipid species. Shaded region represents one standard deviation. (Right) example of curved lipid membrane with compositional asymmetry. Adjacent marker acts as a label as to which series the data represents.}
    \label{fig:compasym}
\end{figure}

With this method, it is also possible to simulate compositionally asymmetric membranes. One extreme example is studying how the adjacent leaflet composition effects the properties of a DPPC monolayer as a function of curvature. To this end, three membranes were simulated at several curvatures. These membranes contained DPPC as the inner and outer leaflet, DPPC outer leaflet with DOPC inner leaflet, and DPPC inner leaflet with DOPC outer leaflet. This resulted in four distinct DPPC monolayers with different neighboring leaflets. Representative snapshots are shown in Fig. \ref{fig:compasym}. In these simulations, the number of lipids is equivalent on each leaflet but the leaflets' lipid type is not.

In Fig. \ref{fig:compasym}, the monolayer thickness is reported as a function of curvature for each of the four DPPC monolayers. 
For all tested membranes, a DPPC leaflet with an adjacent DOPC leaflet is thinner than a purely DPPC membrane. 
It is known that DOPC is a more fluid and thinner leaflet. 
Likewise, DOPC packs less densely than DPPC. 
Thus, due to the equal area constraint, DOPC applies a tension to the DPPC leaflet and thins the membrane. 
However, at high curvatures, this difference appears to become smaller, potentially because bending and stretching behavior can allow the DPPC leaflet to approach its thermodynamically favorable state. While further investigations are necessary,  the applicability of the frozen patches method to study asymmetric curved lipid membranes is demonstrated by these simulations.

\section{Conclusion}

Effective methods for simulating membrane curvature in particle based lipid systems are difficult to construct due to the necessary presence of periodic boundary conditions. Simulating an entire sphere or vesicle with atomic or molecular resolution and explicit solvent is computationally costly. Likewise, any method to simulate membrane curvature without simulating the entire vesicle imposes certain constraints to membrane shape, dynamics, or deformations. Here, we have demonstrated a simple method to impose arbitrary curvature to a membrane while minimizing the artifacts introduced to the bulk of the membrane. This allowed modeling membrane and individual leaflet properties as a function imposed curvature.  Two regimes can be identified, one where the inner and outer leaflet exhibit equal changes in opposite directions, and a second one, where the outer leaflet thins while the inner leaflet remains stagnant. When modeling colloidal lipid nanoparticles and certain organelles, which have radii of curvature  as small as $10-15$ nm \cite{huotari2011endosome,hurley2010membrane,manella2006structure}, the curvatures reach the second ``non-linear'' regime, where the we have shown that the thickness mismatch between leaflets can be as large as $0.3$ nm and the local order parameter can vary up to $0.2$. 

Depending on the system, certain mechanisms which relax the curvature-induced stress can be activated. For instance, a microfluidic synthesis of lipid nanoparticles could self-assemble with a given compositional asymmetry on the leaflets. Additionally, lipids exhibit flip-flop to resolve the differential stress (with high energetic penalty). Cells can use enzymes to maintain this compositional asymmetry. Because of the close relationship between bending and number asymmetry, we investigated membranes that are both curved and contain different number of lipids on each leaflet. As expected, curvature shifted the zero-stress membrane from a symmetric distribution, i.e., with identical number of lipids in each leaflet, to an asymmetric one, with an asymmetry of about 0.2. Interestingly, we observed rich morphology behavior when analyzing membranes with drastic number asymmetries, exhibiting both protrusion and faceting (in a curved membrane) and gel phase transition (in a flat membrane).

The simplicity of this frozen patches method, where pre-equilibrated membrane patches are simply not integrated forward in time, allows the exploration of many open research questions regarding asymmetric and curved membranes. The method created for this work can also easily be extended towards biaxial curvature by freezing a perimeter. This implies that the method presented here can be readily translated to studying the membrane fusion of two spherical caps of lipid membranes. Therefore, commonly expensive simulations of two full vesicles could be then reduced in volume by an order of magnitude by only simulating a fragment of the sphere. The positions of the  membrane patches could also be controlled computationally to model the influence of membrane fluctuation or deformation on these fusion related properties. Finally, we anticipate this method could be useful in simulation of cubosomes (cubically periodic lipid nanoparticles), where frozen patches could enforce a lower boundary of a cubic periodic structure with a super-cell constructed above it.

\section{Appendix}
\subsection{Geometric method for thickness}
\label{app:geometric}
The volume of a cylindrical segment is given by
\begin{equation}
    V = \frac{l_y}{2} \left(\theta \left(r + t\right)^2 - \theta r^2 \right)
\end{equation}
and the surface area of that segment is given by 
\begin{equation} 
A = l_y \left(2t + r \theta + \left( r + t \right) \theta \right)\quad . \label{eq:area}
\end{equation}
Thus, the thickness can be defined in terms of only the area and volume of the semicircle,
\begin{equation} 
t = \frac{1}{4} \left(\frac{A}{l_y} - \sqrt{\left(\frac{A}{l_y}\right)^2 - \frac{16V}{l_y}}\right)\quad . 
\end{equation}
And the average ``radius", given equation \ref{eq:area}, can also be defined,
\begin{equation}
    r = \frac{1}{2 \theta}\left(\frac{A}{l_y}- t\theta - 2t^2\right) \quad .
\end{equation}

\subsection{Curve-fitting fluctuating lipid membranes}
\label{app:curve-fit}
To fit the cylindrically curved but fluctuating lipid membrane with a single curve, the following method was used.
First, we define the imposed curvature as a cylinder with radius $r$,
\begin{equation}
    z_{circ} = \sqrt{\left(r^2-x^2\right)}\quad .
\end{equation}
This equation then is fit separately from the following equations which model the fluctuations atop the membrane. The fluctuations are modeled with the odd sinusoidal terms ($z_{sin}$ with a period of the arc length of the cylinder segment 
\begin{equation}
    z_{sin} = a_1\sin\left(\frac{\pi x}{r\theta}\right)+ \frac{a_3}{3}\sin\left(\frac{3\pi x}{r\theta}\right)+ \frac{a_5}{5}\sin\left(\frac{5\pi x}{r\theta}\right) \quad, 
\end{equation}
as well as the even cosine terms $z_{cos}$
\begin{equation}
    z_{cos} = b_2\cos\left(\frac{2\pi x}{r\theta}\right) + b_4 \cos\left(\frac{4\pi x}{r\theta}\right) \quad .
\end{equation}
To impose the fixed boundary conditions (the membrane can not change in angle nor position at the frozen point), the fit parameters for $z_{sin}$ are constrained by
\begin{equation}
    a_1 + a_3 + a_5 = 0
\end{equation}
and $z_{cos}$ is constrained with 
\begin{equation}
    b_4 = -b_2 \quad .
\end{equation}
Finally, to model the non-circular rearrangement of the membrane (which is needed to ensure good fits), one final cosine fit term was added
\begin{equation}
    z_{adj} = b_1\cos\left(\frac{\pi x}{r\theta}\right) \quad. 
\end{equation} 
The total function which predicts the $z$ coordinate of the data point given the $x$ coordinate is 
\begin{equation}
    z_{tot} = z_{circ} + z_{cos} + z_{sin} + z_{adj} \quad .
\end{equation}
This means the average curvature is fit with one variable, radius, and the fluctuations are fit with an additional 4 independent variables. The an example of the fits is highlighted in Fig. \ref{fig:thumbtacks}.

\section{Supporting Information}
Additional methods validation, analyzed quantities, snapshots, and comparisons to existing methods are included in the supporting information.

\section{Acknowledgments}
We would like to acknowledge Dr. Teshani Kumarage and Prof. Cecilia Leal for insightful conversation about curved lipid membranes and apparent bending moduli. 
This work made use of the Illinois Campus Cluster, a computing resource that is operated by the Illinois Campus Cluster Program (ICCP) in conjunction with the National Center for Supercomputing Applications (NCSA) and which is supported by funds from UIUC.

\bibliography{bib}
\end{document}

\maketitle      
\tableofcontents

\section{Frozen Edges do not Introduce Membrane Defects}

\begin{figure}
    \centering
    \includegraphics[width=.5\linewidth]{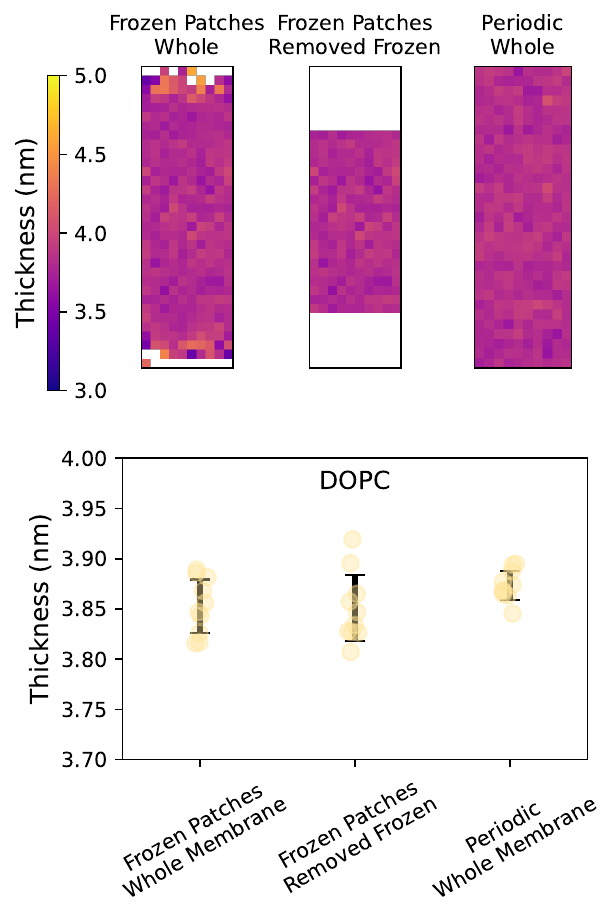}
    \caption{Effect of frozen membrane anchors on equilibrium thickness of DOPC membranes. (Top) Average thickness of frozen anchored DOPC membranes, frozen anchored membrane only counting the ``middle" region, and periodic membrane, averaged over 60 frames. (Bottom) Average thickness of ten independently initialized simulations, averaged over 60 frames.}
    \label{fig:DOPC_flat_thickness}
\end{figure}

In addition to the thickness data presented in the main text, we additionally present the thickness for DOPC membranes. In Fig \ref{fig:DOPC_flat_thickness} to further emphasize that the frozen edges method does not significantly change the mean thickness when sampled across many identical replicates in flat membranes. However, by freezing the membranes, there is an increased standard deviation of the different thicknesses. Again, we see that the properties are shared throughout the whole membrane, and the edges do not appear differently than the bulk of the membrane.
\subsection{Additional Thickness Data for DOPC}

\subsection{Order Parameter Comparison for DOPC and DPPC}
\begin{figure}
    \centering
    \includegraphics[width=.5\linewidth]{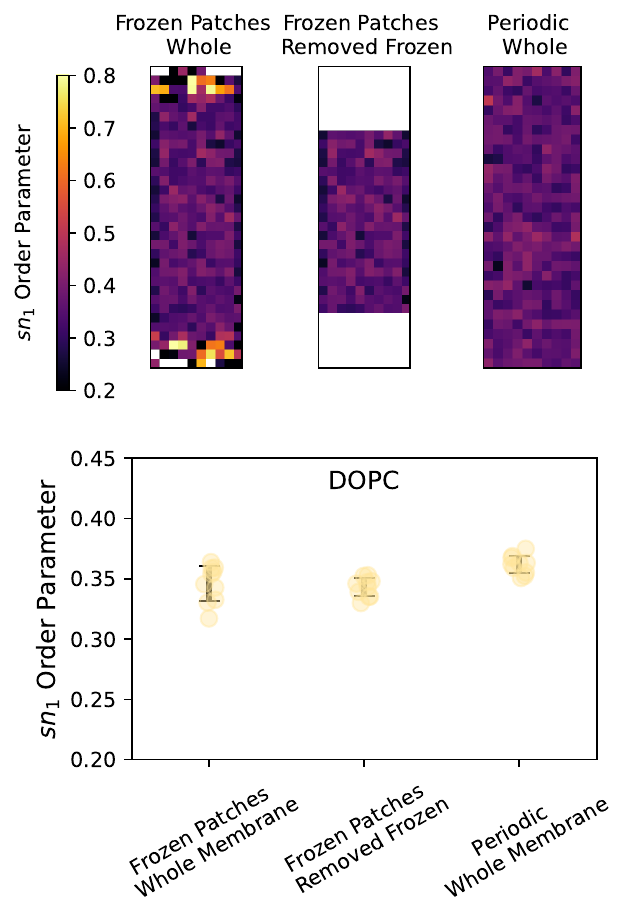}
    \caption{Effect of frozen membrane anchors on equilibrium ordering of DOPC membranes. (Top) Average $sn_1$ order parameter of frozen anchored DOPC membranes, frozen anchored membrane only counting the ``middle" region, and periodic membrane, averaged over 60 frames. (Bottom) Average order parameter of ten independently initialized simulations, averaged over 60 frames.}
    \label{fig:DOPC_flat_order}
\end{figure}

\begin{figure}
    \centering
    \includegraphics[width=.5\linewidth]{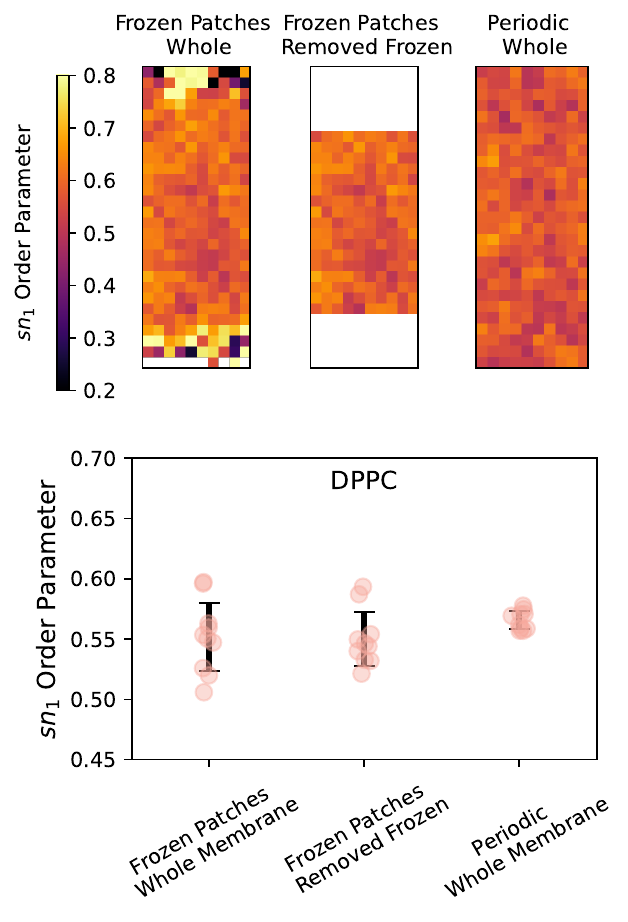}
    \caption{Effect of frozen membrane anchors on equilibrium ordering of DPPC membranes. (Top) Average $sn_1$ order parameter of frozen anchored DPPC membranes, frozen anchored membrane only counting the ``middle" region, and periodic membrane, averaged over 60 frames. (Bottom) Average order parameter of ten independently initialized simulations, averaged over 60 frames.}
    \label{fig:DPPC_flat_order}
\end{figure}

To further emphasize that the frozen edges do not effect the structuring of the membranes, we also demonstrate the order parameter on average in membranes simulated over periodic boundary conditions as well as those simulated with frozen edges. We show the order parameters for DOPC and DPPC membranes in Figs \ref{fig:DOPC_flat_order} and \ref{fig:DPPC_flat_order}

The same trends that can be interpreted from the thickness plots can be interpreted from these order parameter plots.
\subsection{Mean Squared Displacement of flat membrane dynamics}

To assess the dynamics of the membrane, we simulated a membrane with frozen edges for 0.02 $\mu s$ (1,000,000 timesteps). We then calculated the mean squared displacement for the N4a bead in each lipid in the simulation. We colored each lipid by the absolute value of the $x$ position of the N4a bead in the first time step. Beads with larger $|x|$ positions (lighter colors) are closer to the frozen edges. 
\begin{figure}
    \centering
    \includegraphics[width=.5\linewidth]{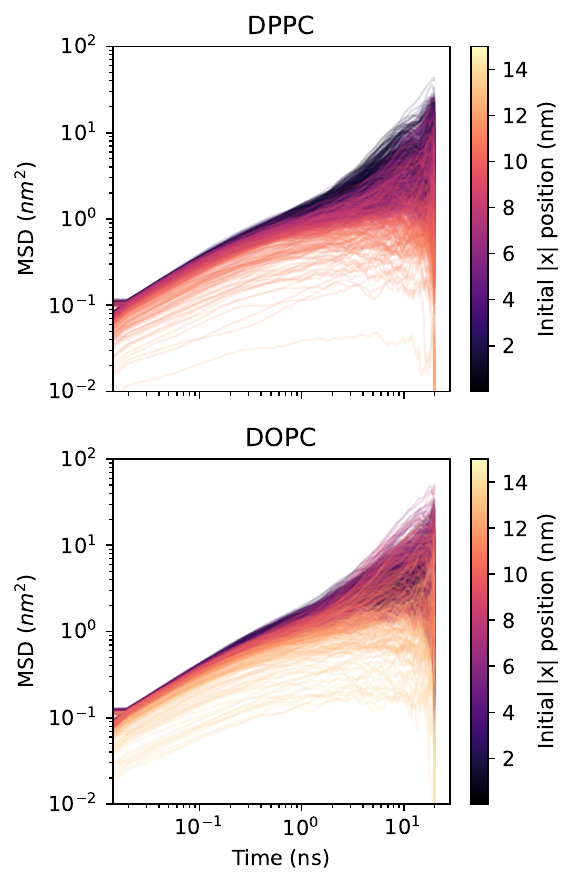}
    \caption{Mean squared displacement of N4a bead for each lipid in one pinned membrane simulation, calculated over 10,000,000 time-steps. Color indicates the absolute value of the $|x|$ position of the lipid membrane in the initial frame, where the maximum value (light color) represents the frozen anchored beads.}
    \label{fig:MSD_test}
\end{figure}

In Fig \ref{fig:MSD_test}, we show these mean squared displacements. In DPPC, it is very clear that lipids with more central positions (dark colors) diffuse faster than other lipids. In DOPC, this trend is less clear, partly because DOPC diffuses faster than DPPC.

\FloatBarrier 

\section{Extracting Curved Membrane Shape}
In addition to Fig. 1 in the main manuscript, which shows the fit curves on top of a rendered snapshot, we show several other curve-fits for certain membrane curvatures and lipid types. 

\begin{figure}
    \centering
    \includegraphics[width=.8\linewidth]{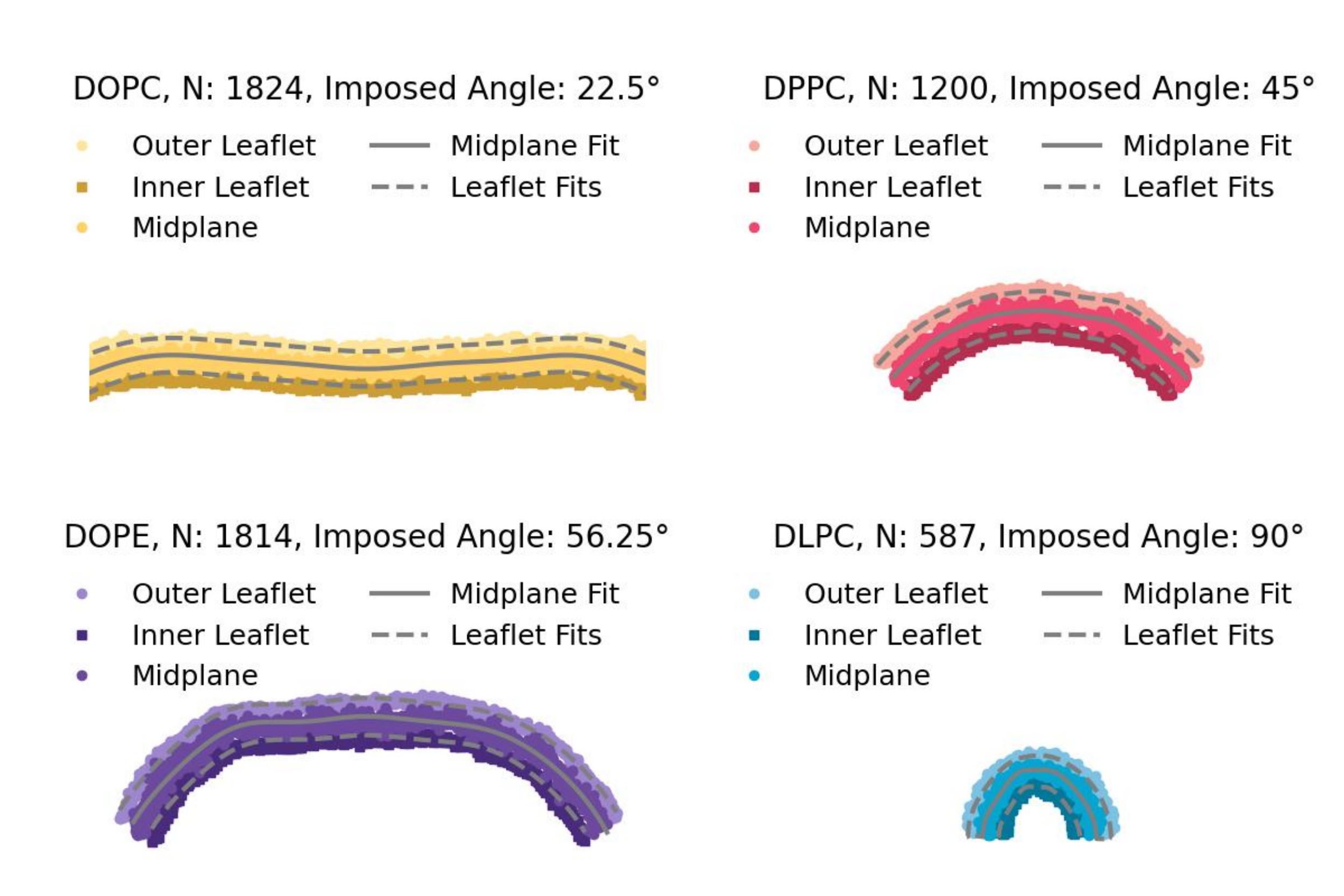}
    \caption{Example 2D data points of curved lipid membranes, varying the lipid type and degree of imposed curvature. Overlayed are the fits for the upper leaflet, lower leaflet, and membrane mid-plane. }
    \label{fig:selected_fits}
\end{figure}

In Fig. \ref{fig:selected_fits}, we show four of these snapshots, highlighting the separation of the outer and inner leaflets (using DBScan) as well as the curve fits for the top, bottom, and membrane midplanes. The fit nicely captures non-circular features and handles different lipid types and membrane sizes with no additional input parameters.
\subsection{Selected fits and Separation of Points with DBScan}

\subsection{Raw Fluctuation Spectra}

\begin{figure}
    \centering
    \includegraphics[width=.7\linewidth]{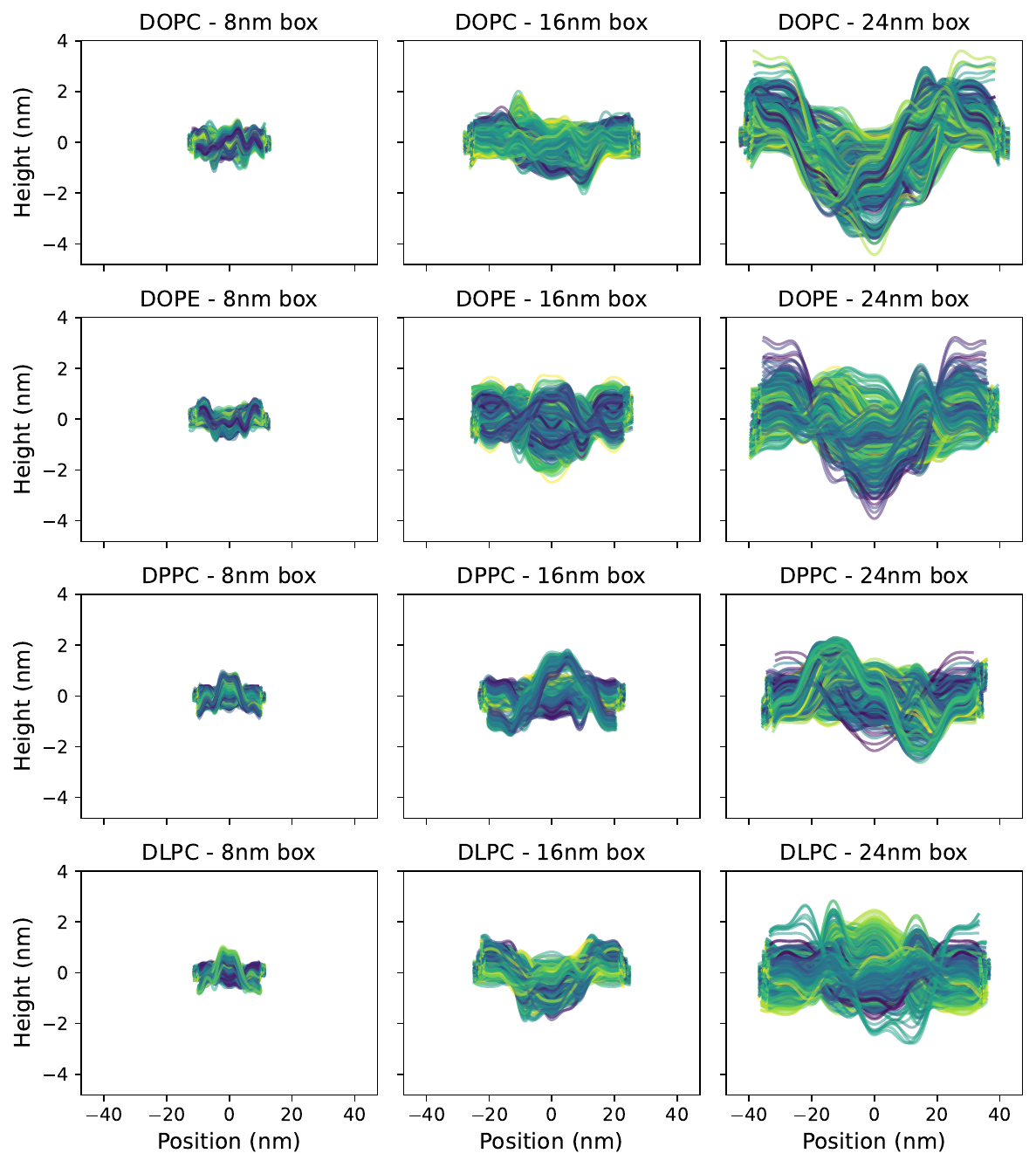}
    \caption{Demonstration of fluctuations (circular component removed) of the curved lipid membranes. Color indicates the radius of curvature of a curve fit. Left to right shows the increasing fluctuation amplitudes with increasing box size. Bottom to top shows the four different lipid types.}
    \label{fig:fluct_raw}
\end{figure}

By removing the semicircle part of the curve fit, we can extract the fits for the fluctuations that live on the average semicircle. In Fig \ref{fig:fluct_raw}, we show these fluctuations for all box sizes, lipids, and curvatures. It is apparent that fluctuations are bigger at higher box sizes. It is also apparent that membranes fluctuate more in the center as compared to the pinned edges.
\subsection{RMS Fluctuation for all Finite sizes}
\begin{figure}
    \centering
    \includegraphics[width=.3\linewidth]{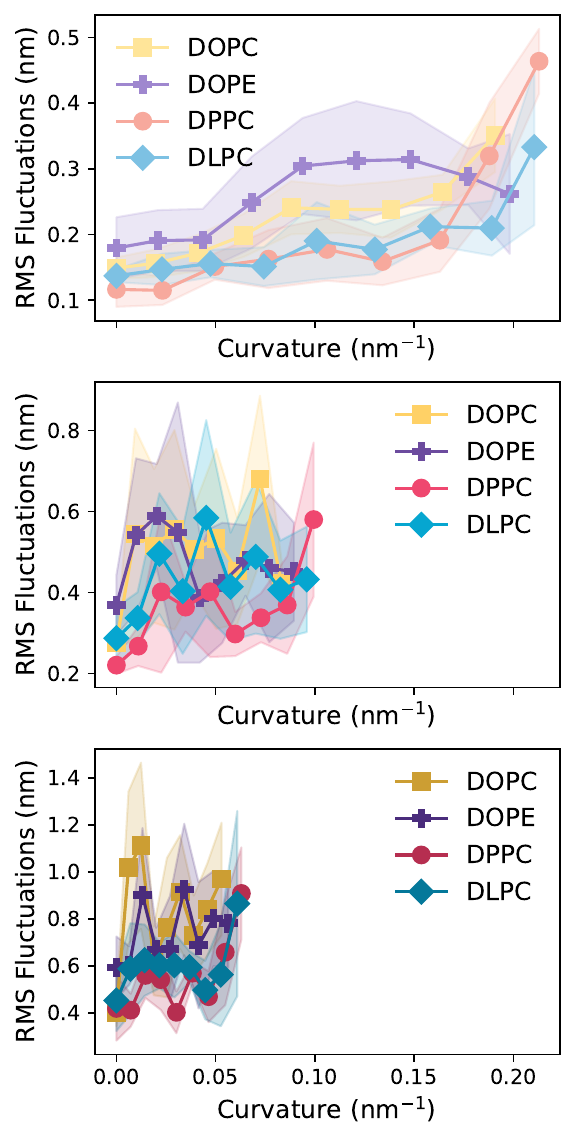}
    \caption{Measure of membrane fluctuation as a function of imposed curvature. Fluctuations are reported as the second moment of the fluctuations which exist on the curved membrane structure. Membranes are reported with 3 separate number of lipids ($N \sim 600, 1200, 1800$), values are averaged over 80 frames and error are represented as the standard deviation of five uniquely initialized snapshots.}
    \label{fig:fluct_all}
\end{figure}

To compliment the main text figure which shows root mean squared fluctuations, we also add the root mean squared of the fluctuations of all number of lipids simulated in Fig \ref{fig:fluct_all}. These show the same ordering unsaturated above saturated lipids.

\FloatBarrier 

\section{Number asymmetry}
\subsection{Additional snapshots of asymmetric thumb-tacked membranes}

\begin{figure}
    \centering
    \includegraphics[width=.35\linewidth]{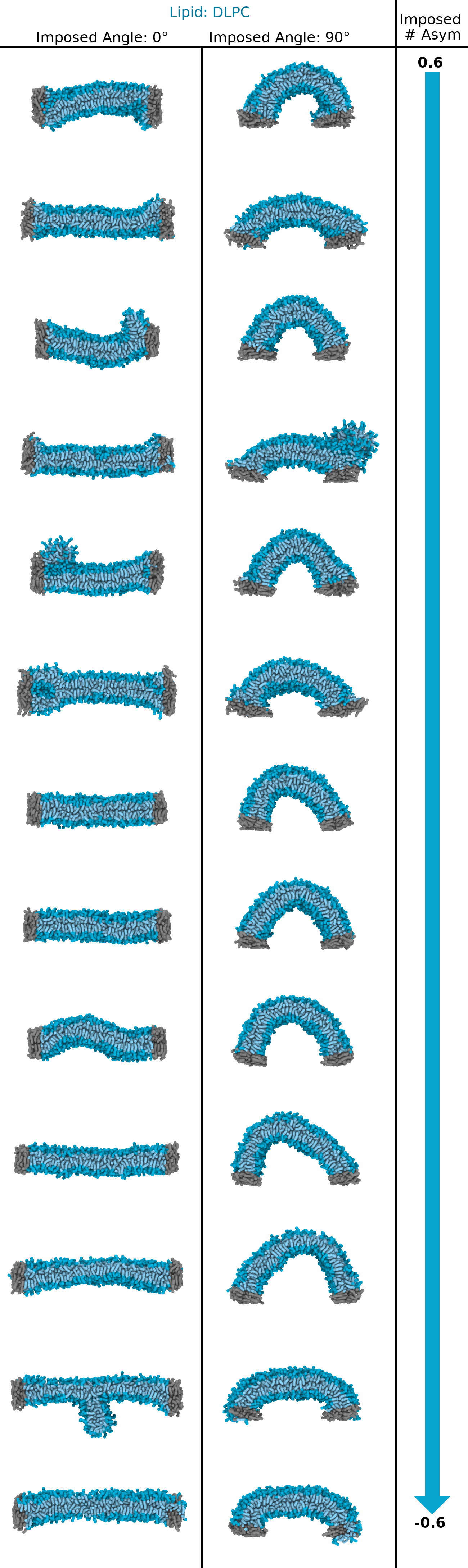}
    \hspace*{5em}
        \includegraphics[width=.35\linewidth]{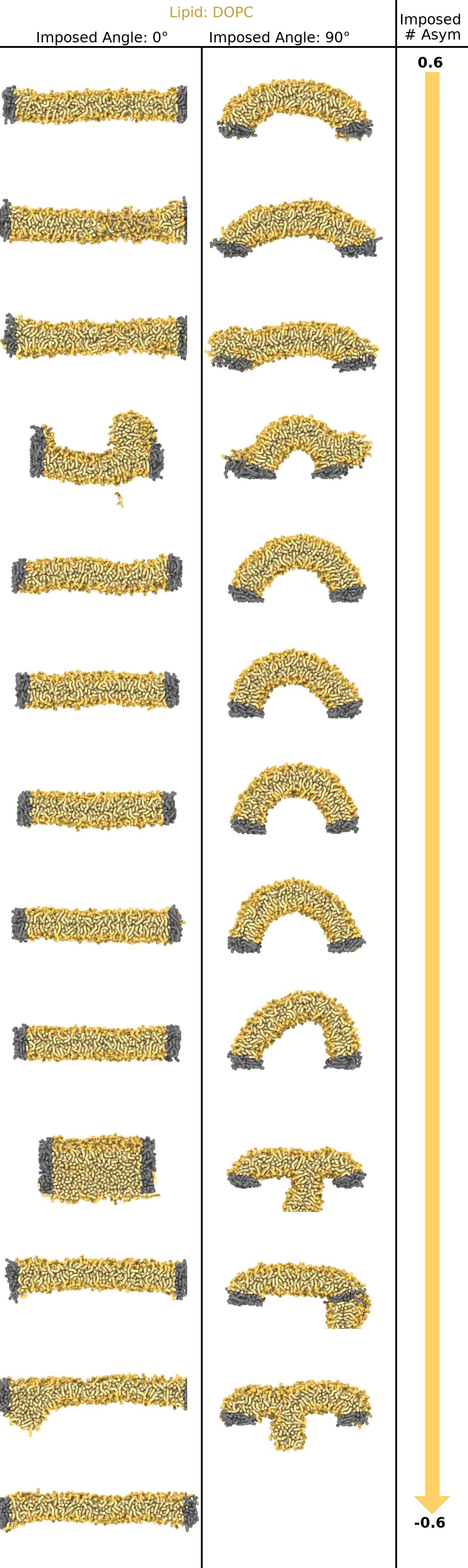}
    \caption{Renderings of DLPC (left) and DOPC (right) membranes initialized with some increasing number asymmetry. Top: overpacked top leaflet. Bottom: overpacked bottom leaflet. Left: no curvature imposed. Right: curvature imposed. }
    \label{fig:megafig_DLPC}
\end{figure}

\begin{figure}
    \centering
    \includegraphics[width=.35\linewidth]{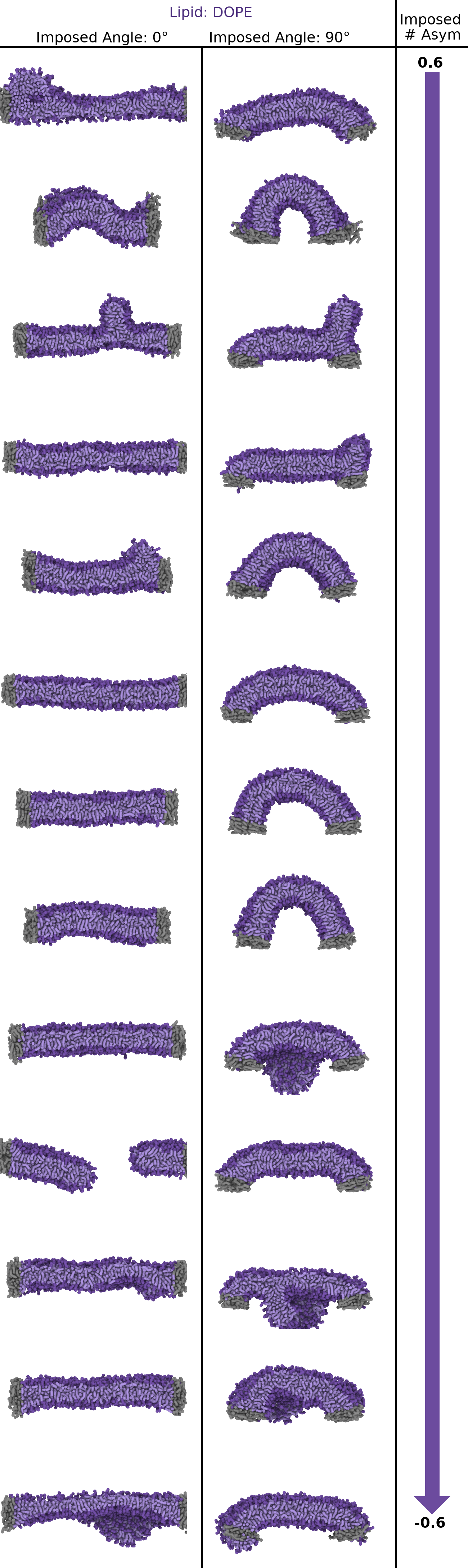}
    \hspace*{5em}
    \includegraphics[width=.35\linewidth]{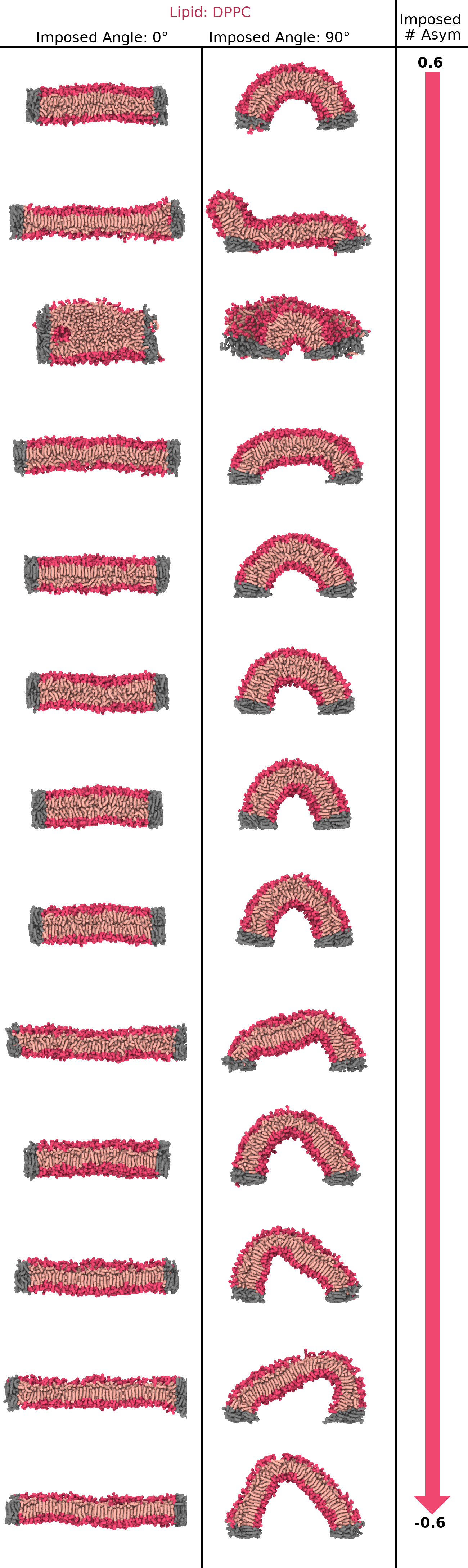}
    \caption{Renderings of DOPE (left) and DPPC (right) membranes initialized with some increasing number asymmetry. Top: overpacked top leaflet. Bottom: overpacked bottom leaflet. Left: no curvature imposed. Right: curvature imposed.}
    \label{fig:megafig_DOPE}
\end{figure}

To compliment the simulations shown in the main text which characterize number asymmetric DPPC membranes, we add snapshots of membranes at a wider variety of number asymmetries from four separate lipid types, shown in Figs. \ref{fig:megafig_DLPC} and \ref{fig:megafig_DOPE}.

Only DPPC shows the transition from a fluid phase to a gel phase under curved conditions. Floppier lipids tend to form more protrusions or otherwise non-membrane structures. At high number asymmetries, the equilibrated patches themselves have defects and membranes are made with defects.

\subsection{Additional asymmetry  plots}

Additionally, we perform further analysis on the asymmetric membranes to understand the extent to which our method can simulate these systems.
\subsubsection{Area, curvature, and number asymmetry relationship}
\begin{figure}
    \centering
    \includegraphics[width=1\linewidth]{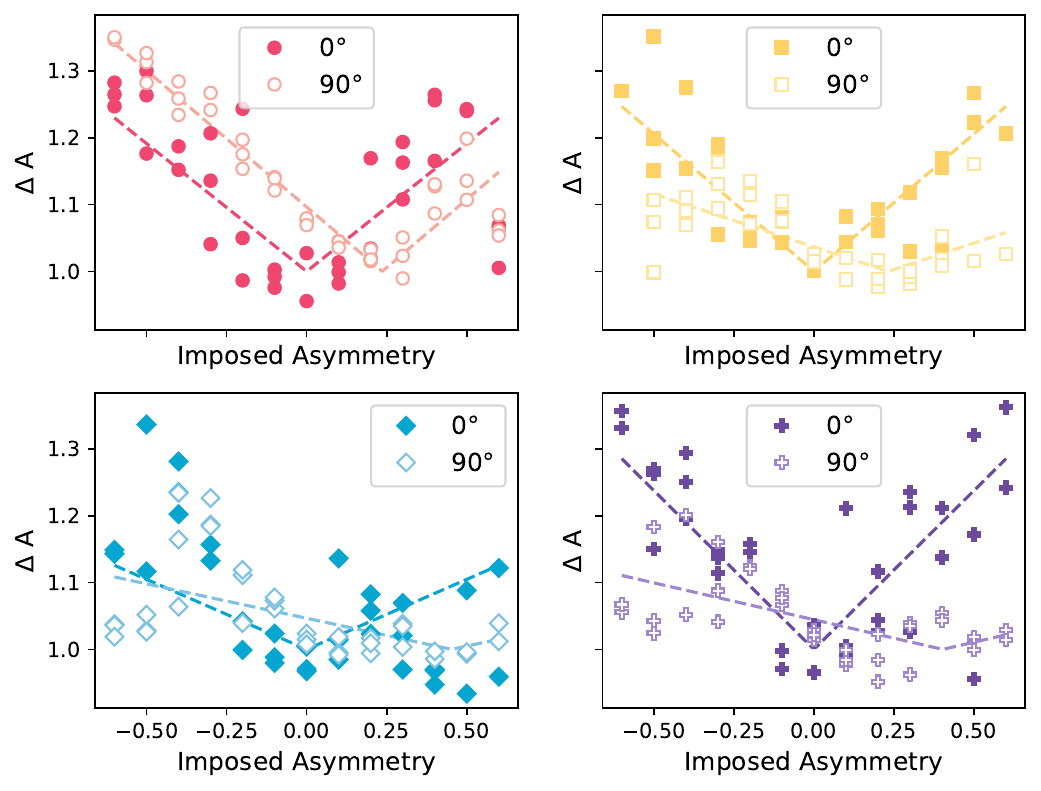}
    \caption{Area of membrane (normalized to minimum membrane area) flat membranes and curved membranes. Lipids DOPC, DPPC, DOPE, and DLPC are tested. Dotted line is a linear fit to the function $\frac{A}{A_0}|n| + k_0$ where $k_0$ represents the spontaneous curvature and $|n|$ represents the number asymmetry.}
    \label{fig:asym_area_all}
\end{figure}

In Fig \ref{fig:asym_area_all}, we show the plot of change in area against imposed asymmetry for all lipid types (similar to the figure shown in the main text). Here, we see the same general trend, that flat membranes have a minimum area at zero asymmetry, and curved membranes have minimum area at some positive imposed asymmetry. It is also noteworthy that for membranes besides DPPC, the data adopts less of a change in area when in the curved state as compared to the flat state. We speculate that this is due to curved membranes being able to relax their area more by undergoing different bending deformations
\subsubsection{Imposed asymmetry and final asymmetry relationship}
\begin{figure}
    \centering
    \includegraphics[width=1\linewidth]{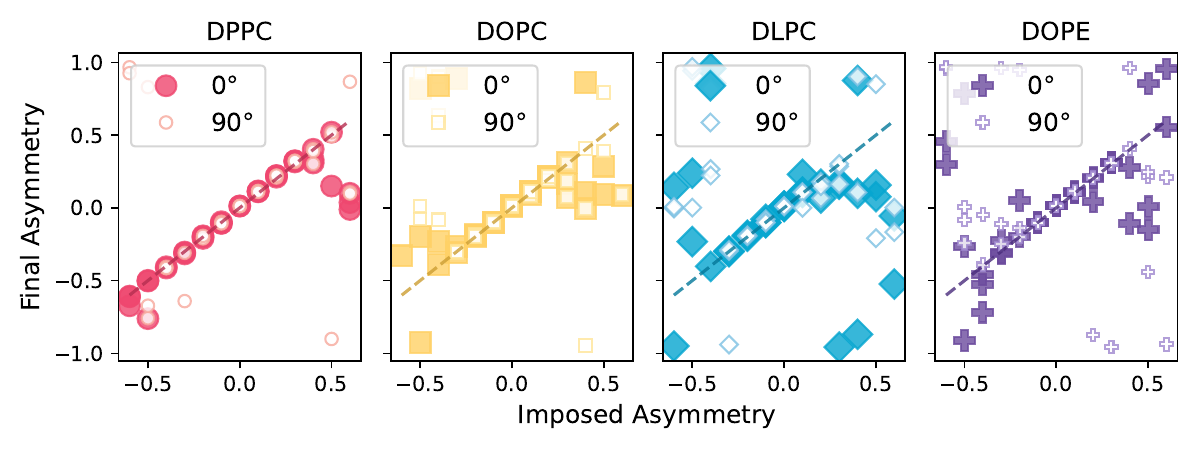}
    \caption{Comparison of initial and final asymmetry of membranes. Initial asymmetry is controlled by where lipids are placed in simulation. Final asymmetry is calculated using HDBSCAN. Line with slope of 1 and no y intercept is shown as a guide for the eye.}
    \label{fig:finalasym}
\end{figure}

Using HDBSCAN, we are able to seperate the top leaflet from the bottom leaflet. We are then able to see how the asymmetry changes after $0.2 \mu s$, as shown in Fig \ref{fig:finalasym}. For each lipid type, there exists a region where the imposed and final asymmetries are the same, and there exists a region where HDBSCAN falters and the final asymmetry measurement no longer matches the imposed asymmetry, likely due to pore formation, protrusion formation, or other erroneous structures. These results qualitatively match the expectation from the snapshots shown above.

Throughout this work, we describe that we do not see lipid flip-flop on the time scale of our simulation. This figure is our strongest evidence of this claim, where despite 10,000,000 timesteps (10 times longer than the symmetric membranes), the final asymmetry is identical to the initial asymmetry under a rather large range of differential stresses. Likewise, curved membranes exhibit similar behavior to flat membranes.

\subsubsection{Imposed asymmetry, lipid order, and curvature relationship}
\begin{figure}
    \centering
    \includegraphics[width=1\linewidth]{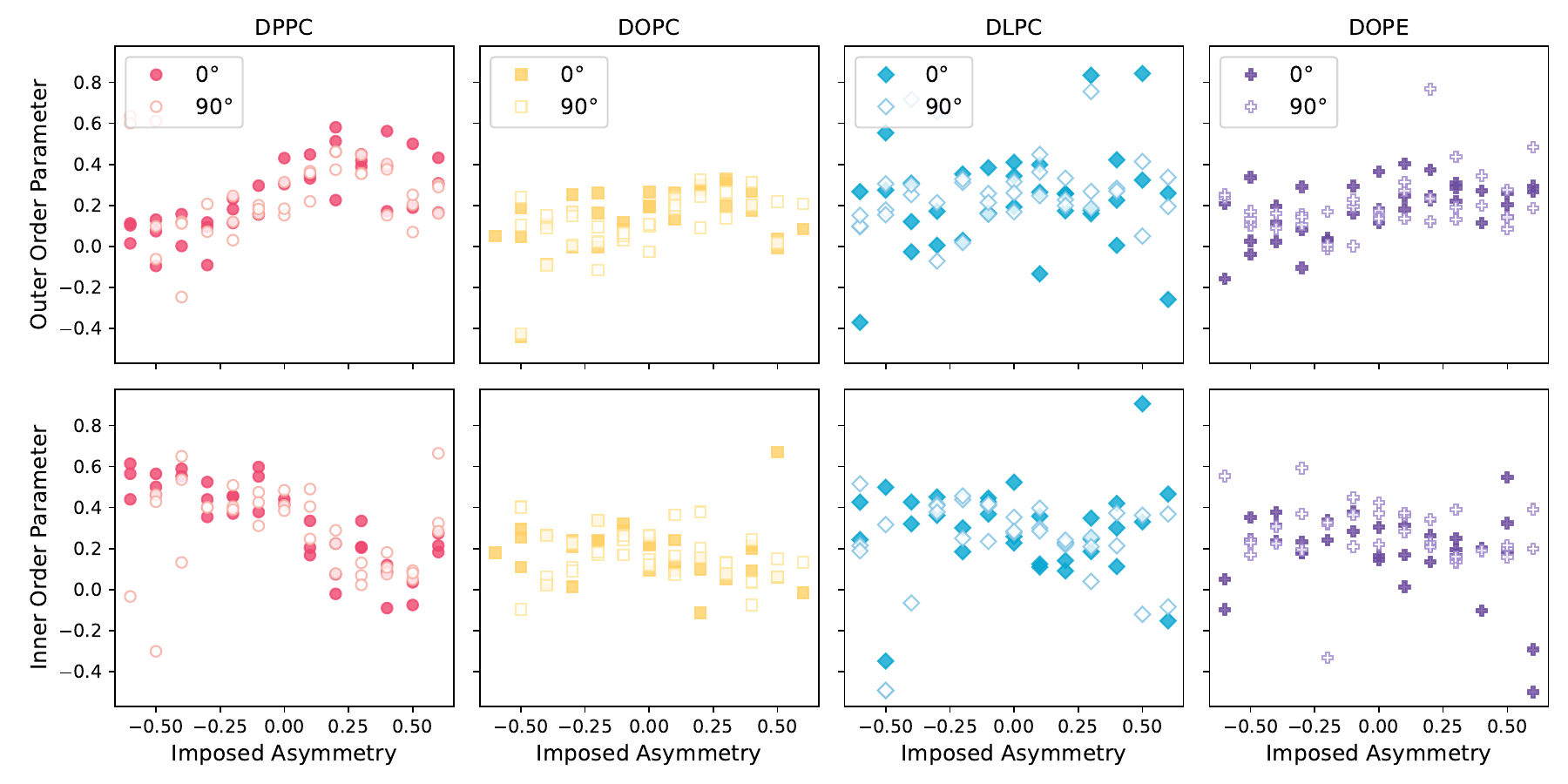}
    \caption{Comparison of order parameter and initial asymmetry of membranes. Initial asymmetry is controlled by where lipids are placed in simulation. Order parameter is calculated using local order parameter with leaflet assignment using HDBSCAN.}
    \label{fig:order}
\end{figure}

Finally, we show the relationship between imposed asymmetry, curvature, and order parameter for the inner and outer leaflets, as assessed with the local curvature method (fig \ref{fig:order}). The order of DPPC was correlated to imposed asymmetry, whereas the other more fluid lipids did not display a clear trend. 

\subsection{Comparison to whole cylinder morphology}

\begin{figure}
    \centering
    \includegraphics[width=0.4\linewidth]{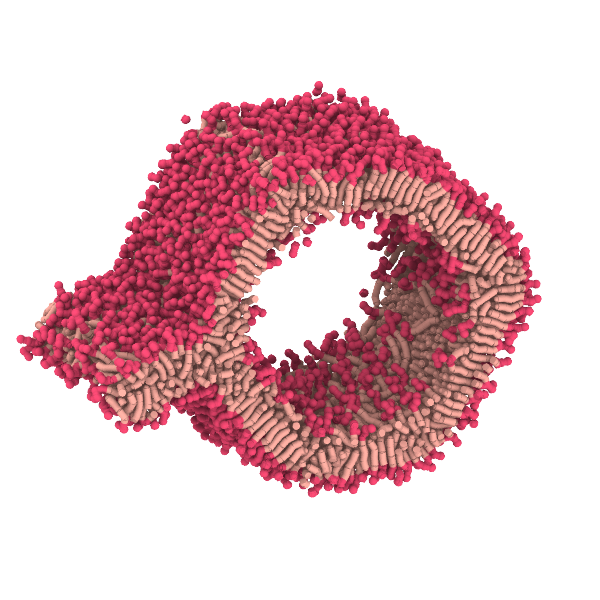}
    \caption{Example snapshot of DPPC membrane initialized in a tube configuration showing similar faceting and protrusions as seen in our thumb tacked membranes.}
    \label{fig:tube}
\end{figure}

In comparison to the work by \citet{Sreekumari2022}, we also show the simulation of cylindrical tubes with number asymmetry and the Martini lipid model. In highly asymmetric systems, protrusions were formed, similar to the structures in the curved lipid membranes. We show one of these protrusions in Fig \ref{fig:tube}. Additionally, slight faceting of the cylinder occurred, in agreement with the faceting observed in the curved membranes(see main text). 

 \FloatBarrier
 
\bibliography{bib}